\documentstyle[11pt]{article}
\title{ Pseudo-Goldstone Bosons in Technicolor Models and the Phenomenology.
\thanks{Invited talk Given at CCAST Seminar of Precision Test of
the SM and Superhigh Energy Physics, Beijing, P.R.China, April 18 to
May 17, 1995;  to appear in the Proceedings.} }
\author{ Zhenjun Xiao$^{a,b}$ \thanks{Email: lugr@bepc2.ihep.ac.cn} and
Xuelei Wang$^{a,b}$ \\
{\small a. CCAST(World Laboratory), P.O.Box 8730, Beijing 100080,
P.R.China}\\
{\small b. Department of Physics, Henan Normal University,}\\
{\small Xinxiang, Henan, 453002 P.R.China.\thanks{Mailing Address} }\\ }
\date{}

\begin{document}
\maketitle
%%%%%%%%Put up preprint number
\begin{picture}(0,0)(0,0)
\put(310,205){{\large HNU-TH/95-16}}
\put(310,190){{\large July 5, 1995}}
\end{picture}
%%%%%%%%%%%%%%%%%%
\begin{abstract}
In this report we present a review of recent developments
in the TC/ETC theories, concentrating on the theoretical
estimations and the phenomenological analysis  about the
Non-Oblique corrections on the $Zb\overline{b}$ vertex
from ETC dynamics and Pseudo-Goldstone Bosons.
The relevant studies about the vertex corrections
on other processes from the PGBs were also considered.
\end{abstract}

\newcommand{\beq}{\begin{eqnarray}}
\newcommand{\eeq}{\end{eqnarray}}
%%%% the widths
\newcommand{\zbb}{Zb\overline{b}}
\newcommand{\gu}{\Gamma_u}
\newcommand{\gd}{\Gamma_d}
\newcommand{\gnu}{\Gamma_\nu}
\newcommand{\gb}{\Gamma_b}
\newcommand{\gh}{\Gamma_h}
\newcommand{\gz}{\Gamma_Z}
%%%%%%% the ratios
\newcommand{\rb}{R_b}
\newcommand{\rc}{R_c}
\newcommand{\rl}{R_l}
%%%%%%%% the vertex factors
\newcommand{\dbnew}{\Delta_{bv}^{new}}
\newcommand{\dbpa}{\Delta_{bv}^{P^{\pm}}}
\newcommand{\dbpb}{\Delta_{bv}^{P_8^{\pm}}}
\newcommand{\dbetc}{\Delta_{bv}^{ETC}}
\newcommand{\sx}{(S, T, U, V, W, X)}
\newcommand{\mh}{M_H}
\newcommand{\mt}{m_t}
\newcommand{\dpi}{\delta \Pi}
\newcommand{\dwpi}{\delta \widehat{\Pi}}
\newcommand{\zbs}{Z \rightarrow b\overline{s}(\overline{b}s)}

\tableofcontents
%\newpage

\section{ Introduction}

\hspace{0.5cm}
After the discovery of top quark and the measurement of its mass at
Fermilab \cite{CDF,D0} the investigation for the mechanism
of the electroweak symmetry breaking(ESB) becomes the number one task
facing particle physics society.
At present there is no any evidence to show which mechanism is
responsible for the ESB.
The fundamental Higgs boson, which is
responsible for the spontaneous symmetry breaking in the Standard
Model(SM) \cite{Glashow},
has not been discovered so far despite the intensive searching in
experiments. This lack of experimental observation of the elementary
Higgs boson is one of the main motivations for constructing models of
dynamical electroweak symmetry breaking(DESB).

Since last Spring, it has been widely reported that there is a discrepancy
between the measured $R_b$, $R_b=\Gamma(Z\rightarrow b\overline{b})/
\Gamma(Z\rightarrow hadrons)$, at LEP and the theoretical prediction
of the SM:
\beq
R_b^{exp} &=& 0.2202\pm 0.0020, \; \hbox{(1994, ref.\cite{Schaile2})}
 \nonumber\\
          &=& 0.2204\pm 0.0020, \; \hbox{(1995, ref.\cite{LEP1995})},
\label{eq:rbexp}
\eeq
while in the SM,
\begin{eqnarray}
R_b^{SM}=0.2158\pm 0.0004\ \ for \ \ m_t=180 \pm 12\;GeV
\end{eqnarray}
It is easy to see that the $R_b^{SM}$ is approximately
$2-\sigma$ away from the measured $R_b$. A positive
contribution to $R_b$ from new physics clearly is required to lift
the $R_b$ to the measured value.
Of cause we understand that the size
of measured $R_b$ may decrease along with the progress of the experiments
\cite{Behnke}.
However, if the observed deviation is really the long-awaited deviation
from the SM, its implication for those new theories beyond standard model
are very interesting.

In theories of DESB, such as
technicolor(TC) \cite{Weinberg}, electroweak symmetry breaking
is due to chiral symmetry breaking
in an asymptotically-free, strongly-interacting, gauge theory with massless
fermions.
Technicolor and extended technicolor (ETC)
\cite{Weinberg,Farhi,Dimopoulos}
theories generally yield large effects on the physical observables.
A common approach to studying the new physics effects is to assume that the
dominant effect comes from oblique
corrections, which have been conveniently parametrized in terms of three
parameters S, T and U \cite{Peskin} or $\epsilon_1$, $\epsilon_2$,
$\epsilon_3$ and $\epsilon_b$ \cite{Altarelli}.
More recently, the (S, T, U) parametrization has
been extended by introduction of additional three parameters (V, W, X)
\cite{Burgess1,Evans1,Burgess2}.

In general, contributions from vertex and box diagrams are usually tiny.
But for the $\zbb$  vertex, the situation
is changed greatly \cite{Cornet,Xiao3}.
In fact, the non-oblique corrections on the
$Zb\overline{b}$ vertex from the sideways and diagonal
ETC gauge boson exchanges \cite{Simmons,Chivukula,Kitazawa,Wu,Yue1995},
as well as
from the charged Pseudo-Goldstone bosons (PGBs) exchanges \cite{Xiao1,Xiao2}
could be rather large, which will affect the partial decay width
$\Gamma_b=\Gamma(Z\rightarrow b \overline{b})$ and consequently
the ratio $R_b$ and other relevant observables.  The systematic studies
about non-oblique corrections on the $\zbb$  vertex are very interesting
for one to look for the new physics effects on precisely measured
observables.

In a previous review paper \cite{King}, S.F.King has described a general
picture of the basic structures and recent developments of dynamical
electroweak symmetry breaking.
In this report we concentrate on the studies about the non-oblique
corrections on the $\zbb$  vertex from ETC gauge boson exchanges
(sideways and diagonal) and from
the charged PGBs, and to see that what implications are there for TC/ETC
theories, if the true values of $m_t$, $\rb$  and other relevant observables
are within their reported $1-\sigma$ error.

This paper is organized as follows: In section 2 we
first list the new data reported at the 1995 Winter conference\cite{LEP1995},
and then present the theoretical predictions for $R_b$ and other
observables in the SM. Discussions about the particle spectrum of
simple TC/ETC  models and possible experimental signatures
are condensed into the section 3.
The original ideas and recent developments about the parametrization
of oblique corrections in TC theory are presented in section 4.
In section 5 we discuss the non-oblique corrections on the $\zbb$  vertex
form various sources, especially from the exchanges of
charged PGBs which appeared
in the QCD-like one-generation TC model(OGTM) \cite{Farhi}.
In section 6 we briefly discuss and comment on
several new TC models proposed very recently in the sense of
avoiding the existed constraints imposed by the precision data.
The conclusions are also included in section 6.

%%%%%%%%%%
%%%%%%%%%%
\section{ The $Zb\overline{b}$ vertex in Standard Model}

\hspace{0.5cm}
With LEP entering into its final period of measurements on the Z peak,
the accuracy achieved in LEP experiments now reaches a very high level,
as illustrated in Table \ref{tab:table1} \cite{LEP1995,Fleming1}.
The current precision achieved at LEP (Moriond 1995) \cite{LEP1995}
and at SLC
experiments now permits very rigorous tests of the SM  and encourage us to
study   possible discrepancies
between experiments and SM predictions. While the SM is generally in
excellent agreement with experiment, recent results on the left-right
asymmetry $A_{LR}$ at SLC \cite{SLC} and the ratio $\rb$ measured at LEP
\cite{Schaile2,LEP1995} indicate a possible disagreement at 2 to $2.5\sigma$
level. The values of strong coupling constant $\alpha_s$, measured at the
low-energy experiments and at the $M_Z$ scale respectively, also show some
disagreement.

\begin{table}[htbp]
\caption{ The experimental values for the precision Z-pole observables,
directly quoted from ref.(25)}\label{tab:table1}
\begin{center}
\vspace{.1cm}
\begin{tabular}{c|c|c} \hline\hline
Quantity & \hspace{.8cm} experimental Value \hspace{.8cm} &
\hspace{.4cm}Standard Model Fit \hspace{.4cm}\\ \hline
$M_Z\;(GeV)$ & $91.1887 \pm 0.0022$ & input\\
$\gz\;(GeV)$ & $2.4971\pm 0.0033$ & 2.4979\\
$\sigma ^h_p\;(nb)$ & $41.492 \pm 0.081$ & 41.441\\
$R_e=\gh/ \Gamma_e$ & $ 20.843 \pm 0.060 $ & 20.783\\
$R_\mu=\gh/ \Gamma_\mu $ & $ 20.805 \pm 0.048 $ & 20.783\\
$R_\tau=\gh/ \Gamma_\tau$ & $ 20.798 \pm 0.066 $ & 20.783\\
$A^0_{FB}(e)$ & $0.0154\pm 0.0030$& 0.0157\\
$A^0_{FB}(\mu)$ & $0.0160\pm 0.0017$& 0.0157\\
$A^0_{FB}(\tau)$ & $0.0209\pm 0.0024$& 0.0157\\
$A_{\tau}(P_\tau)$ & $0.140\pm 0.008$& 0.145\\
$A_e(P_\tau)$ & $0.137\pm 0.009$& 0.145\\
$R_b$ & $0.2204\pm 0.0020$& 0.2157\\
$R_c$ & $0.1606\pm 0.0095$& 0.172\\
$A^0_{FB}(b)$ & $0.1015\pm 0.0036$& 0.1015\\
$A^0_{FB}(c)$ & $0.0760\pm 0.0089$& 0.0724\\
$A^0_{LR}$ & $0.1637\pm 0.0075$& 0.145\\ \hline \hline
\end{tabular}
\end{center}
\end{table}

In this report we concentrated on the investigations about the ratio $\rb$,
for relevant studies  of other two observables the reader can see
the refs.\cite{Bamert,Shifman}

%%%%%%%%%%%%%%%%%%%%%%
\subsection{The top quark and Higgs boson}
\hspace{0.5cm}
In last Spring, the CDF collaboration first
published \cite{CDF2} the evidence for the existence of the top quark.
In this March, the CDF \cite{CDF} and D0 \cite{D0} collaborations at Fermilab
announced the observation of top quark in $p\overline p$ collisions at the
Tevatron. Both groups saw a statistically significant excess of dilepton
and lepton + jet events with the proper kinematic properties and bottom
quark tags needed to indicate $t\overline t$
production. Furthermore, they were
able to extract mass values of top quark by fitting to events consisting
of a single lepton plus four jets.  The CDF group found that
$m_t=176 \pm 8 \pm 10\;GeV$ \cite{CDF}, while D0 Collaboration obtained
a mass of $m_t=199 ^{+19}_{-21}\pm 22\;GeV$ \cite{D0},
and the weighted average is $m_t=180 \pm 12\;GeV$.
The measured top quark mass is in very good agreement with the prediction
based on the SM electroweak fits of the LEP and other data,
$m_t=178 \pm 11 ^{+18}_{-19}\;GeV$ \cite{Schaile2}, where the
central value and the first error refer to $M_H=300\;GeV$.
This measurement of $m_t$, while
still not very precise, should help in reducing the present uncertainties
on almost all electroweak observables.
The direct observation of the top quark at the Tevatron heralds the start of
a new era in the study of particle physics.

The top quark is certainly unique among the ordinary fermions. It is the
heaviest fermion discovered so far, more than 30 times as massive as the
bottom quark. Correspondingly, top quark has the
largest coupling to the symmetry
breaking sector of all the known particles. This large coupling to the Higgs
sector may give rise to deviations from its expected behavior,
thereby offering  clues to electroweak symmetry breaking, fermion mass
generation, quark family replication, and other deficiencies of the standard
model. Obviously, the knowledge of
$m_t$ will be very helpful for one to look for the hints of new physics.
According to current estimation
the combined CDF+D0 determination of $m_t$ could provide
an overall error $\Delta m_t= \pm 3 \;GeV$ \cite{Mangano} by the
end of this century.
It will probably be necessary to wait for an NLC to get
$m_t$ with an accuracy of less than 1 $GeV$ \cite{Treille1994}.
For a general discussion of top quark physics one can see the paper
written by C.P.Yuan \cite{Yuan1995}.

After the discovery of the top quark and the measurement of its mass the main
uncertainties of the SM expectations for those observables are clearly due to
our ignorance about the Higgs boson mass.
Theoretically, Higgs boson mass is a free parameter of the SM.
If Higgs bosons exist as discernible states, theoretical consistency
demands that they lie below about $700-800\;GeV$. The current lower
limit is $M_H\; > \;63\;GeV$ \cite{Wyatt}, which is coming from the failer
of the direct searches at LEP. The LEP 200 can rise this limit to about
$90\; GeV$.
On the other hand, the steadily increasing accuracy of the data
starts to exhibit some  weak
sensitivity to the Higgs boson mass \cite{Ellis1}.
As described in ref.\cite{Langacker} the $\chi^2$ distribution generally
predicts
a light Higgs boson. However, the constraint is weak statistically. From
the $\chi^2$ distribution one can obtains the weak upper limits
\cite{Langacker}
\beq
M_H< 510 (730)\; GeV \ \  \hbox{at 90(95)\%C.L},
\eeq
from the indirect precision data and the CDF measurement of $m_t$ (where
$m_t=174\pm 16\; GeV$ was used). This sensitivity to the $M_H$
is driven almost entirely by the measured $R_b$ and $A_{LR}^0$, both of which
are well above the corresponding SM expectations. Omitting these two data
leads to an almost flat $\chi^2$ distribution, as illustrated in Fig.11 of
ref.\cite{Langacker}. If the present deviations of $R_b$ and $A^0_{LR}$
are due to  large statistical fluctuations or new physics beyond the SM the
above upper constraints on $M_H$ will disappear.

Although some researchers claim that the current data prefer a
relatively light Higgs boson \cite{Ellis1}, but in fact there is no definite
constraint on $\mh$ existed now. A recent analysis done by M.Consoli
and Z.Hoiki \cite{Consoli1} using the data from the 1995 Winter conference
suggest that the Higgs boson
mass should be in the heavy range, say, $M_H \sim 500 -
1000\; GeV$ for $m_t=180\;GeV$. This result is in agreement with the
indications from the $\Delta r$ $(e.g., \ \ M_W$) analysis \cite{Xiao5}.

Further improvement in the data taking is needed for a definite
answer to the value of $M_H$. Consequently, it is dangerous
to focus on a light-mass region in Higgs searches at future experiments.
According to the studies in refs.\cite{Consoli1,Consoli2} it is possible
to obtain precious information on the Higgs mass when the top quark
mass will be measured with a high precision.

%%%%%%%%%%%%%%%%%%%%%%
\subsection{Non-oblique corrections on the $\zbb$  vertex in SM }

\hspace{0.5cm}
For LEP processes there are two types of radiative corrections:
the corrections to the gauge boson self-energies and the corrections
to the $\zbb$ vertex. In  the evaluation
of self-energy  corrections the error due
to  our ignorance of the Higgs mass
is substantial after the direct
measurement of $m_t$ at Fermilab \cite{CDF,D0}.
On the other hand, in the corrections to the $\zbb$ vertex,
where the leading contribution due to the large top quark mass is
produced by the exchange of the W bosons, there is no dependence on the
unknown Higgs mass. Moreover, the possible new physics contributions
to the $\zbb$ vertex are much more restricted.
Any non-standard behavior most
possibly means the existence of new physics!

The observables (which are in close relation with the
$\zbb$ vertex) considered in this paper include
$\gb$, $\gh$, $\gz$, $\rb$, $R_c$
and $R_l$,
they are well determined theoretically and experimentally.
Because the asymmetry $A_{FB}^b$ is almost
unaffected by the $\zbb$ vertex correction
\cite{Altarelli} we will not include this quantity in our analysis.

In the framework of standard model
calculations of the one-loop corrections to the $\zbb$ vertex
has been performed by several groups
\cite{Akhundov}.
The partial decay width $\Gamma (Z\rightarrow f \overline{f})$ has been
calculated in the $\overline{MS}$ renormalization scheme
\cite{Degrassi}
and has been expressed in a compact form
\cite{Pich},
\beq
 \Gamma(Z\rightarrow f\overline{f})&=& \frac{N_c^f}{48}\frac{\hat{\alpha}}
{\hat{s}^2_w \hat{c}_w^2}\,m_Z
[\hat{a}_f^2 + \hat{v}_f^2](1+\delta^{(0)}_f)(1+\delta_{QED}^f)\nonumber \\
&&\cdot (1+\delta_{QCD}) (1+\delta_\mu^f)(1+\delta_{tQCD}^f)(1+\delta_b),
\label{eq:zff}
\eeq
where $N_c^f=3(1)$ for quarks (leptons) is the color factor, $\hat{\alpha}$
is the electromagnetic coupling constant defined at the $M_Z$ scale,
$\hat{s}_w^2$ is the Weinberg angle in the $\overline{MS}$ scheme, and the
$\hat{v}_f$ and $\hat{a}_f$ are the effective vector and axial
coupling constants of the Z boson to the fermions.
The partial decay widths in eq.(\ref{eq:zff})
has included the genuine electroweak
corrections, the QED and QCD corrections, as well as the corrections to
$\zbb$ vertex due to the large top quark mass.
The definitions and the explicit expressions for all functions and
factors appeared in eq.(\ref{eq:zff}) can be
found in refs.\cite{Degrassi,Pich}.
In ref.\cite{Fleischer},
J.Fleischer {\it et al}  calculated the two-loop $0(\alpha\alpha_s)$
QCD corrections to the partial decay width $\gb$, and they found a
screening of the  leading one-loop top mass effects by $m_t\rightarrow $
$m_t\,[1-\frac{1}{3}(\pi^2-3)\alpha_s/\pi]$.
The expression for $\Gamma(Z\rightarrow f\bar{f})$ in eq.(\ref{eq:zff})
is very convenient for the calculation of branching ratios because most
factors will be canceled in the ratios of widths.
For more details about the calculations  of $\Gamma_i$ and other
relevant quantities in the SM one can see
the refs.\cite{Akhundov,Degrassi} and a more recent paper \cite{Xiao3}.

In our analysis, the measured values
\cite{CDF,D0,Schaile2,LEP1995,pdg,BES}
$m_Z=91.1887$ $\pm 0.0022\;GeV$, $G_\mu=1.16639\times 10^{-5}(GeV)^{-2}$ ,
$\alpha^{-1}=137.0359895$, $\alpha_s(m_Z)=0.125\pm0.005$,
$m_e=0.511\;MeV$, $m_\mu=105.6584\;MeV$ and $m_\tau = 1776.9\;MeV$,
together with $m_t=180 \pm 12 \;GeV$ and the assumed value
$M_H=300^{+700}_{-240}\;GeV$ are used as the input parameters.
In the  numerical calculations
we conservatively take the ``on-shell'' mass of the
b-quark the value $m_b=4.6\pm 0.3\;GeV$ (in ref.\cite{Pich},
the authors used
$m_b=4.6\pm 0.1\;GeV$), and use the known relation \cite{ckg} between
the ``on-shell'' and the $\overline{MS}$ schemes to compute the running mass
$\overline{m_b}(m_Z)$ at the Z scale.
We also use the same treatment for the c-quark and take $m_c=1.5\,GeV$ as
its ``on-shell'' mass. For other three light quarks we simply assume that
$\overline{m_i}(m_Z)=0.1\,GeV\;(i=u,d,s)$.
All these input parameters will
be referred to as the {\em Standard Input Parameters} (SIP).

Using the SIP, the SM predictions for Z decay widths
and the ratios can be calculated easily.
The size of uncertainties in $\Gamma_i$ and $R_j$ depend on the
errors of $\mt$, $\mh$. $\overline{m_b}(M_Z)$, $\alpha_s(M_Z)$ and
$\hat{\alpha}$. For instance, the partial decay
width $\gb$ and the ratio $R_j$ (j=b, c, l)
can be written in the following form:
\beq
\Gamma_b &=& 377.8 \pm 0.2(m_t) _{+0.2}^{-0.9}(M_H)
\pm 0.5(\alpha_s)  \pm 0.4(\hat{\alpha})\label{eq:gb}\\
R_b  &=& 0.2158 \pm 0.0004(m_t) \pm 0.00003(M_H)
\pm 0.00004(\alpha_s) \pm 0.0001(\overline{m_b})\label{eq:rbsm2} \\
R_c &=&0.1722\; \pm 0.0002(m_t)\; \pm 0.00004(M_H)
\pm 0.0001(\alpha_s) \pm 0.00003(\overline{m_b})\label{eq:rc} \\
R_l&=&  20.820 \pm 0.002(m_t) ^{-0.015}_{+0.011}(M_H)
\pm 0.034(\alpha_s) \pm 0.003(\overline{m_b}) \label{eq:rl}
\eeq
where the central value corresponds to $m_t=180\;GeV$, $M_H=300\;GeV$,
$\alpha_s(M_Z)=0.125$ and $\overline{m_b}(M_Z)=2.8\;GeV$. The contributions
to the Z boson decay width from the $0(\alpha^2)$ terms are less than
0.1 MeV and can be neglected completely.

Among the electroweak observables
the ratio $R_b$ is the special one \cite{Xiao3}. For this ratio
most of the vacuum polarization corrections
depending on the $m_t$ and $M_H$
cancel out, while the experimental uncertainties
in the detector response to hadronic events also basically cancel.
Furthermore, this ratio is also insensitive to extensions of the SM
which would only contribute to vacuum polarizations.
Analytically, the ratio $R_b$ has a complicated dependence on the $m_t$
and $M_H$ in the region under study ( $m_t=180\pm 12\;GeV$,
$M_H=60\sim 1000\;GeV$ ). Its plot is shown in Fig.1.
The two parameter ($m_t$ and $M_H$) fitting to the
exact results gives
\beq
R_b^{SM}&=& 0.21892 -10^{-4}\cdot\left [ 7.45 \frac{m_t^2}{m_Z^2}
+1.75 \ln \left[\frac{m_t^2}{m_Z^2}\right]
- 0.098 \ln \left[ \frac{M_H^2}{m_Z^2}\right] \right].
\label{eq:rbfit}
\eeq
The errors introduced with this parametrization in the region
under study is completely negligible (less than $0.00001$)

For the ratio $R_c$, the current accuracy of the data is limited by
systematic effects, coming from the large bottom contamination in the
charm samples \cite{Brown}. The new LEP value $R_c=0.1606 \pm 0.0095$ is in
good agreement with the SM prediction although the error of the data is
still large.

The current accuracy of the ratio $R_l$ is very high, $R_l=20.820\pm0.035$
\footnote{$R_l=20.820 \pm 0.035$ is the weighted average
of the measured $R_e$, $R_\mu$ and $R_\tau$ as given in Table 1.},
the relative error is
about $0.17\%$. In SM, the $R_l$ is practically a constant for given $m_Z$
due to (accidental) cancellations between the universal and vertex
contributions. Non-standard terms would spoil this cancellation and exhibit
a deviation from the SM value. However, the present data does not show
deviations from the SM.

For the ratio $R_b$, the situation is more interesting:

$\bullet$ \hspace{.5cm} The direct measurement of $m_t$ at CDF and D0,
$m_t=180 \pm 12\; GeV$, while
still not very precise, has provided a great help in reducing
the theoretical uncertainty of $R_b$ to 0.0004.

$\bullet$ \hspace{.5cm} The two-loop $0(\alpha \alpha_s)$
QCD contribute a 0.0006 positive correction to
the central value of  $R_b$.
As illustrated in Fig.1, where the lower curve with solid triangle symbols
 shows the $R_b$ at one-loop level and the upper line
with solid square symbols represents the $R_b$ with the
inclusion of $0(\alpha \alpha_s)$ QCD corrections,
the two-loop QCD contribution makes the $R_b$ moving in the right
direction toward the range preferred by the data.

$\bullet$ \hspace{.5cm}  From the Fig.1, it is easy to see that the
$R_b$ is about two standard deviations away from the central value
of the measured $R_b$.
The deviation reaches 2.2-$\sigma$ (or 2.5-$\sigma$ at
one-loop order) for $m_t=180\;GeV$.

Because of special vertex corrections, the partial width $\Gamma_b$
actually decreases with $m_t$, as opposed to the other widths
which will increase. The ratio
$R_b$ is insensitive to the still unknown Higgs boson mass $M_H$.
However, when combined with other observables, for which $m_t$
and $M_H$ are strongly correlated, the effect is to favor a smaller
Higgs mass, as discussed previously. If the current deviation of $R_b$
is more than a statistical fluctuation, it must be due to some sort of new
physics. We know that many types of new physics,
such as the TC/ETC theories, will couple preferentially to the 3rd
generation, so the careful investigations about the possible
contributions to the $\zbb$ vertex form the new physics
are certainly very important !

%%%%%%%%%%%%%%%%%%%%%
\subsection{ The vertex factor $\Delta_b^{new}$}

\hspace{0.5cm}
The precision data can be used to set limits on TC theory
as well as other kinds of new physics.
Besides the $m_t$
dependence the $\zbb$ vertex is also sensitive to a
number of types of new physics. One can parametrize such effects by
\cite{Langacker}
\begin{eqnarray}
\Gamma_b=\Gamma_b^{SM}(1+\Delta_b^{new})\label{eq:gbdef}
\end{eqnarray}
where the term $\Delta_b^{new}$ represents the pure
non-oblique corrections to the $\zbb$ vertex from new
physics, while the oblique corrections to
$\gb$ have been neglected.
The partial decay width $\Gamma_b^{SM}$ can be determined
theoretically by eq.(\ref{eq:zff}),
and other five observables $(\Gamma_h,\; \Gamma_Z,\; R_b,\; R_c,\; R_l)$
can be written in a general form
\beq
O_i = O_i^{SM} + \lambda_i\cdot \Delta_b^{new}
+ \sum_{j=1}^{6}C_{ij}\cdot P_j,\label{eq:ghdef}
\eeq
Where $P_j$ represent oblique parameters $(S, T, U, V, W, X)$, and
$C_{ij}$ are the corresponding
coefficients respectively.

The parameter W appears in the decay
width of the W boson, but not in the precision electroweak observables
studied here. Concerning X, explicit calculations in ETC models
\cite{Burgess1,Evans1,Burgess2} find that the X parameter is
very small in all
scenarios, so it can also be neglected in our studies about the $\zbb$
vertex. The parameter V may become significant for small technifermion
masses \cite{Burgess2} $M_{N,E,U,D}\leq M_Z$. However, following J.Ellis's
argument, we also regard this possibility as unlikely.
According to recent studies \cite{Langacker} the
parameters $(S, T, U)$ are all close to zero with
small errors. On the other hand,
the oblique corrections
will be basically canceled
in the ratios $(R_b, R_c, R_l)$, the corresponding coefficients
should be very small. In short,
since we here concentrate on estimating the
non-oblique corrections on the $\zbb$ vertex
from new physics and studying its implications for TC/ETC theories,
we could neglect all those six oblique parameters approximately ($e.g.,$
we do one-parameter fit).

The definition of
$\Delta_b^{new}$ in eq.(\ref{eq:gbdef}) is
different from that of $\epsilon_b$
\cite{Altarelli}(as well as the parameter $\Delta_b$ in
refs.\cite{Blondel,Cornet}), and this vertex factor $\dbnew$ represents
the pure non-oblique corrections on the $\zbb$ vertex.

In a previous paper \cite{Xiao4} we used the likelihood function method
to derive out the
value of $\dbnew$ from the data set
$(\Gamma_b, \Gamma_h, \Gamma_Z, R_b, R_c, R_l)$.
With the SIP, the point which maximizes
${\cal L}(x_{exp}, \Delta_b^{new})$ is found to be \cite{Xiao4}
\beq
\Delta_b^{new} &=& 0.017\pm0.007\ \ at\ \ 68\%C.L.,\ \
\hbox{(only $\Gamma_b$ and  $R_b$ included)}, \\
               &=& 0 \pm 0.005\ \ at \ \ 68\%C.L., \ \
\hbox{(all six  observables included)},
\eeq
for $m_t=180\; GeV$ and $M_H=300\;GeV$. One can  also
obtain the $95\%$ one-sided upper (lower) confidence
limits on $\Delta_b^{new}$:
\beq
-0.011 < \Delta_{b,exp}^{new} < 0.011, \ \
\hbox{ (all six observables included)}
\label{eq:dbnewl}
\eeq
for $m_t=180\pm 12\;GeV$ and $60\;GeV\leq M_H \leq 1000\;GeV$.

In the following analysis we will use the $\Delta_{b,exp}^{new}$
as the experimentally determined vertex factor.

%%%%%%%%%%%%%%%%%%%%%%%%%%%
%%%%%%%%%%%%%%%%%%%%%%%%%%%
\section{Pseudo-Goldstone bosons in TC models}

\hspace{0.5cm}
The subject of dynamical electroweak symmetry breaking (DESB) has a long and
distinguished history going back to the early work of Nambu and Jona-Lasinio
 \cite{Nambu}). A series of pioneering papers followed
which extended these ideas to the realm of gauge theories.
Eventually the idea of technicolor (TC) was introduced
by Susskind and Weinberg \cite{Weinberg} as a mechanism for DESB. The early
development of TC is nicely traced in the collection or reprints by
Farhi and Jackiw \cite{Farhi2}.

TC theory is modeled on the known behavior of quarks
in QCD -- but scaled up to the TeV scale.
It turns out that TC by itself is not sufficient to provide
fermion masses. One way forward is to embed the TC gauge group
into a larger gauge group known as ETC.
However we shall see
it is not an easy task to describe the quark and lepton
mass spectrum without running into phenomenological problems.
These problems include the flavor-changing neutral current (FCNC)
problem, problems of producing the correct spectrum for ordinary fermions,
specifically the heavy top quark mass.
These problems have thwarted attempts to construct ETC models,
and to date there is no accepted standard ETC model in the
literature.
Recently ETC has staged a comeback
due to a lot of exciting progress with the above problems.
Such as the invention of ideas of ``Walking TC'' \cite{Holdom}
and ``Strong ETC'' \cite{Evans2}.
Both these ideas result in the technifermion $T$ condensate
receiving a high momentum enhancement,
while the pion decay constants $F_{\pi}$ which depend on
low momentum physics are almost unchanged.
This is important since the quark and lepton masses and PGB masses
depend upon the value of the condensate, while the $W,Z$ masses
depend upon $F_{\pi}$. Condensate enhancement may therefore
increase fermion masses without increasing gauge boson masses.

In this section we focus on the studies about the
spectrum of pseudo-Goldstone bosons in the OGTM.
For recent progress of TC and ETC theories,
the reader can also see the review papers written by Lane \cite{Lane2},
King\cite{King}, and by Chivukula {\it et al} \cite{Chivukula3}.

%%%%%%%%%%%%%%%%%%%%%%%%%%%%%%
\subsection{PGBs in the OGTM }

\hspace{0.5cm}
In this section we consider the single techni-family scenario,
and briefly discuss the resulting pseudo-Goldstone boson phenomenology.
Consider a TC model based on the gauge group,
\beq
SU(N)_{TC}\otimes SU(3)_C \otimes SU(2)_L \otimes U(1)_Y
\eeq
and with a single techni-family,
\beq
\begin{array}{ccl}
{Q_L}^{\alpha} =
\left( \begin{array}{c}U_L \\ D_L \end{array}
\right)^{\alpha} &\sim& (N,3,2,1/6)\\
{U_R}^{\alpha} &\sim& (N,3,1,2/3)\\
{D_R}^{\alpha} &\sim& (N,3,1,-1/3)\\
{L_L}^{\alpha} =
\left( \begin{array}{c}N_{L} \\ E_L \end{array}
\right)^{\alpha} &\sim& (N,1,2,-1/2)\\
{E_R}^{\alpha} &\sim& (N,1,1,1)\\
{N_R}^{\alpha} &\sim& (N,1,1,0)
\end{array}
\eeq
where $\alpha=1 \ldots N$ is the $G_{TC}$ index.
The technifermions which carry technicolor and QCD color are referred
to as techniquarks, while the technifermions which carry technicolor
but not QCD color are called technileptons. Note that the
right-handed technineutrino $N_R$ is required by anomaly
cancellation, and cannot be given a Majorana mass without breaking
$SU(N)_{TC}$.
The ordinary quarks and leptons transform as usual and are technicolor
singlets. In the limit that QCD and electroweak interactions are switched
off, the TC sector of the model respects a large chiral symmetry
$ SU(8)_L\otimes SU(8)_R$. Electroweak symmetry is broken by the
condensate $<\overline{T} T> \ne 0$.
Since the techniquark condensates have QCD color,
there are really eight separate condensates above, which break
the chiral symmetry down to $SU(8)_{L+R}$. The electroweak symmetry
is now broken by the equivalent of four separate technidoublets
(e.g., $N_D=4$)
and thus the gauge boson masses are given by
\beq
M_W = \frac{1}{2} g (\sqrt{4} F_\pi),\ \
M_Z= \frac{g \sqrt{4} F_\pi}{2 \cos_{\theta_w}},
\eeq
and the technipion decay constant mass is now,
\beq
F_{\pi}=\frac{246}{\sqrt{4}} GeV=123 GeV
\eeq
According to Goldstone's theorem we would expect $8^2-1=63$ massless
(Pseudo)-Goldstone bosons produced from this breaking,
one for each broken generator.
Three of them are eaten  by
the Higgs mechanism, while the remainder are assumed to get masses from
a combination of color, electroweak gauge interactions and ETC
interactions.

The 60 PGBs of the $SU(N)_{TC}$ model
consist of the following states,
\beq
P_8^{\pm} &\sim&  \overline{Q}\gamma_5\lambda_{\alpha}\tau^1 Q
\pm i\overline{Q}\gamma_5\lambda_{\alpha}\tau^2 Q \nonumber \\
P_8^0 &\sim&  \overline{Q}\gamma_5\lambda_{\alpha}\tau^3 Q\; ,   \ \
P_8^{0'} \sim \overline{Q}\gamma_5\lambda_{\alpha}Q \nonumber \\
T^a_i &\sim& \overline{Q_i}\gamma_5\tau^a L, \ \ \
T_i \sim \overline{Q_i}\gamma_5L \nonumber \\
\overline{T}^a_i &\sim& \overline{L}\gamma_5\tau^a Q_i,\ \ \
\overline{T}_i \sim \overline{L}\gamma_5Q_i \nonumber \\
P^{\pm} &\sim&  \overline{Q}\gamma_5\tau^{\pm}Q
- 3\overline{L}\gamma_5\tau^{\pm}L
\nonumber \\
P^{0} &\sim& \overline{Q}\gamma_5\tau^{3}Q - 3\overline{L}\gamma_5\tau^{3}L
\; , \ \
P^{0'} \sim  \overline{Q}\gamma_5Q - 3\overline{L}\gamma_5L
\eeq
where $Q=(U,D),\; L=(N,E)$,
$\lambda^{\alpha}$ ($\alpha=1 \ldots 8$) are the Gell-Mann
color matrices and $\tau^a$ ($a=1 \ldots 3$) are the Pauli
isospin matrices.
These 60 PGBs can be classified as follows:

$\bullet$ \hspace{.5cm} The four color octets, which form an isotriplet,
$P_8^\pm $
with charges $\pm 1$ and $P_8^{0}$,
and an isosinglet, $P_8^{0'}$.

$\bullet$ \hspace{0.5cm} Four color triplets and four color anti-triplets,
which form one  isotriplet $T^a_i$ and its self-conjugate
$\overline{T}^a_i$,
and  isosinglet $ T_i$ and its conjugate $\overline{T}_i$.
They are composites made
out of a techniquark and a technilepton or {\it vice versa}. We usually
refer to them as leptonquark PGBs.

$\bullet$ \hspace{.5cm} Four color singlet, which also form a
triplet ($P^{\pm}$ and
$ P^0 $) and singlet $P^{0'}$ of isospin.

In the OGTM, besides the presence of PGBs, the vector resonances
($\rho_T$'s and $\omega_T$) will also appear. All these particles may be
classified by their $SU(3)_c$ and $SU(2)_V$ quantum numbers as shown in
Table \ref{pgbs}.

\begin{table}[htbp]
\begin{center}
\caption{Spectrum of particles in non-minimal technicolor models.}
\label{pgbs}
\vskip.5pc
\begin{tabular}{|c|c|c|c|}   \hline
$ SU(3)_C$    &$SU(2)_{V}$
&PGBs & V-resonances\\ \hline
$1$      &$1$  &$P^{0 \prime}$ & $ \omega_T$  \\ \hline
$1$      &$3$  &$P^{0,\pm}$ & $ \rho^{0,\pm}_T$  \\ \hline
$3$      &$1$  &$P^{0 \prime}_{3}$ & $ \rho^{0 \prime }_{T 3}$  \\ \hline
$3$      &$3$  &$P^{0,\pm}_{3}$  & $\rho^{0,\pm}_{T 3}$  \\ \hline
$8$      &$1$  &$P^{0 \prime}_{8} (\eta_T) $  &
$ \rho^{0 \prime }_{T 8}$  \\
\hline
$8$      &$3$    &$P^{0,\pm}_{8} $  & $ \rho^{0,\pm}_{T 8}$  \\ \hline
\end{tabular}
\end{center}
\end{table}

%%%%%%%%%%%%%%%%%%%%%%%%%%%%%%%%%
\subsection{Estimation for the Masses of PGBs}

\hspace{0.5cm}
There are several kinds of contributions to the masses of PGBs:
electroweak interactions, QCD interactions, and ETC
interactions.
Peskin and Preskill \cite{Peskin2} have calculated the contributions to the
PGB masses due to the color and electroweak interactions. The ETC
contribution to these masses have been worked out by Bin$\acute{e}$truy
{\it et al}  \cite{Binetruy}.

In the limit where standard model interactions and ETC
interactions are turned off, the PGBs would be massless.
Turning on gauge interactions causes  the  PGBs to receive mass
contributions from graphs with a single gauge boson exchange.
At first the electroweak contributions to the masses of PGBs are
theoretically well understood and can be reliably computed (with
some dependence on the TC model) \cite{Peskin2}:
\beq
M_{P^{\pm}}|_{EW} \approx 5 - 14\; GeV.
\eeq

For colored PGBs the QCD contributions to their mass will be dominant.
For $SU(N_{TC})$ TC models with QCD-like dynamics one can estimate
the QCD contributions to the colored PGBs \cite{Peskin2}:
\beq
M^2 |_{QCD} = 3\alpha_s M_{TC}^2  \approx  3\alpha_s
\left [ \frac{8F_{\pi}}{\sqrt{N_{TC}}} \right ]^2.
\eeq
where $F_{\pi}=246/\sqrt{N_D}\; GeV$ is the TC analog of the QCD $f_\pi$.

With the inclusion of electroweak and QCD contributions one can obtains
the masses of PGBs in the SU scenario as follows \cite{Peskin2}:
\beq
Color\ \  singlets,\ \  P^\pm, P^0, P^{0 \prime} &&  \sim 10\;GeV\nonumber\\
Color\ \  triplets,\ \   &&  160-170\sqrt{4/N_{TC}}\;GeV\nonumber\\
Color\ \  octets,\ \  P_8^\pm, P_8^0, P_8^{0 \prime} &&
246\sqrt{4/N_{TC}}\;GeV\label{eq:masspgbs}
\eeq
These masses could be increased in walking TC theories since
the condensate enhancement also enhances PGB masses.
We also expect additional uncertainties for
models (multiscale, strong ETC)
where the TC dynamics is quite different from QCD.

Finally, turning on the ETC interactions can give rise to masses
for the PGBs. Although the ETC contributions to the
PGB masses are entirely  model dependent, one expects, based on Dashen's
formula \cite{Dashen}, that these contributions have the
following form \cite{Chivukula3}:
\beq
M_P^2|_{ETC}  \approx {{<{\overline \Psi } \Psi {\overline \Psi } \Psi
>}\over{F_{\pi}^2 \Lambda_f^2}},
\eeq
where $\Psi$ is the technifermion field,
and $\Lambda_f\equiv M_{ETC}/g_{ETC}$ is the
ETC scale associated with an ordinary fermion $f$.  Assuming that the vev
of the four-fermion operator factorizes, one have:
\beq
M_P|_{ETC} \approx {{<{\overline \Psi } \Psi >}\over{F \Lambda_f}}
\approx {{m_f}\over{F }}\,\Lambda_f.
\eeq

Further more, if there exists a consistent
dynamical model of EWSB which can produce
a heavy enough $t$ quark, then using a $t$ quark mass of 180 GeV,
$F_{\pi}$ of 123 GeV, and an ETC scale at least as large as the technicolor
scale of a TeV, we have a contribution to the PGB mass of the order
of 1 TeV!   Thus it may not be surprising if
PGBs are not found at colliders any time soon.
Consequently,  the masses of PGBs will be considered as
``free'' parameters in the following phenomenological analysis.

%%%%%%%%%%%%%%%%%%%%%%%%%%%
\subsection{Possible experimental signatures of PGBs}

\hspace{0.5cm}
The experimental signatures of the PGBs were studied by Ellis {\it et al}
\cite{Ellis1981}, Dimopoulos \cite{Dimopoulos1980},
and Eichten {\it et al}
\cite{Eichten1986}. Very recently, Chivukula {\it et al}
\cite{Chivukula3} presented a long report to summarize
the possible signatures of colored PGBs and resonances at existing
and proposal colliders.

We here only provide a scanning of possible experimental signatures
of the PGBs. For more details see the papers mentioned above.

It is relatively straightforward to find the color octet
PGBs at the LHC, but much harder (but not impossible) at the Tevatron.
Consider the color octet neutral state $P_8^{0 \prime}$, which
is a techni-isospin singlet and can be produced
singly in hadronic collisions with a cross-section
$d\sigma /dy \approx 1 (10^{-2})$ nb at the LHC (Tevatron)
(for rapidity $y=0$).
The $P_8^{0 \prime}$ can decay back into $gg$ or into $t\overline{t}$ if
kinematically allowed. The first signal at the Tevatron may be
an enhancement of the top quark
production cross-section, as discussed by Eichten {\it et al}
\cite{Eichten1986} and Appelquist and Triantaphyllou \cite{Appel1992}.
In fact the CDF and D0 cross-section for $t\overline{t}$ production
does appear to be slightly higher than
standard model expectations \cite{CDF,D0,CDF2}
\footnote{Last year, CDF published the evidence of top quark
production with $\sigma_t=13.9^{+6.1}_{-4.8}\; pb$ \cite{CDF2};
In this March, both
CDF and D0 announced the discovery of top quark with
$\sigma_t=6.8^{+3.6}_{-2.4}\; pb$ \cite{CDF}, and
$\sigma_t=6.4\pm 2.2\; pb$ \cite{D0}; while the SM prediction is
$\sigma_t^{SM}=4.5\pm 0.3\; pb$ for $m_t\approx 180\;GeV$.
Although the central value of the measured $\sigma_t$ is larger than that
predicted by the SM, but the measured and theoretically predicted top quark
production cross-section are obviously agree within $1-\sigma$ level.}.

The color triplets are examples of leptonquarks. There are many of
them in the spectrum, consisting of color triplet combinations such as
$U\overline{N}$, $U\overline{E}$, $D\overline{N}$, $D\overline{E}$ and
their antiparticles. At the LHC or HERA they
are copiously pair produced and tend to decay
into heavy quarks and leptons with relatively background-free signatures.
For example a typical signature of a leptonquark pair might be
$t\overline{t}\tau \overline{\tau}$ which has a particularly low background.
For more details about
the productions and decays of leptonquarks the reader can see a new
report written by A.Djouadi {\it et al} \cite{Djouadi1995}.

The color singlets are similar to charged and neutral Higgs bosons.
The best place to look for them is the clean environment provided
by the high energy $e^+ e^-$ colliders, although they should also
be seen at the LHC.
The neutral PGBs do not have a tree-level coupling
to the $Z$ boson, however, which should enable it to be
distinguished from neutral Higgs bosons.
They couple to gauge bosons via the triangle anomaly,
with technifermions running round the loop,
as discussed in some detail in ref.\cite{Ellis1981,Chivukula3}.

In ref.\cite{Lubicz1995}, Lubicz and Santorelli estimated the production
and decay of neutral PGBs at LEP II and NLC in multiscale walking
technicolor (WTC) models. They found that, in Lane-Ramana multiscale
model \cite{Lane1991}, because of
the existence of relatively low TC scales, the production of neutral
PGBs, in $e^+ e^-$ colliders LEP II or NLC, is significantly enhanced.
This enhancement is expected to increase the corresponding cross sections
by one or two orders of magnitude with respect to the prediction of
traditional TC models. The neutral PGBs could be observed mainly
in the processes
$e^+ e^- \to P \gamma$, $P e^+ e^-$ or $P Z^0$ (if kinematically allowed)
at LEP II and at NLC.

The charged PGBs couple to the photon and $Z$ by tree-level
couplings which resemble those for charged Higgs,
and like charged Higgs tend to decay into the heaviest fermions around.
The current lower limit on the mass of charged Higgs bosons is generally
equivalent to the lower limit on the mass of color singlet PGBs, $e.g.,$
$M(P^\pm) > 41.7\;GeV$ at present \cite{ALEPH1992}.

Finally note that the colored PGBs may re-scatter into
eaten technipions, and hence may enhance the rates of longitudinal
gauge boson scattering, as observed by Bagger, Dawson and Valencia
\cite{Bagger1991}.

In order to give an overlook for the discovery potential of those PGBs
appeared in TC models we
present the Table \ref{signature}
(directly quoted  from ref.\cite{Chivukula3}),
which summarize the discovery reach of different machines.
For a more general study about the searchers for new particles
at existing and proposal high energy colliders one can see
ref.\cite{Treille1994}.

\begin{table}[htbp]
\begin{center}
\caption{Discovery reach of different accelerators for particles associated
with realistic models of a strong EWSB sector, the masses in GeV.
Directly quoted from ref.(55) with small modification.}
\label{signature}
\vskip 0.5pc
\begin{tabular}{|c||c||c||c||c||c|}\hline\hline
           &          &      &   &     &    \\
Particle   &Tevatron  &LHC   &LEP I &LEP II  &TLC  \\  \hline\hline
           &          &      &        &   &  \\
$P^{0 \prime}$  &---    &$110 - 150$  $^a$   &$8$  $^{b}$ ;
$28$  $^{b}$  &---$^c$  &---$^c$                   \\ \hline
           &          &      &        &  &  \\
$P^{0 }$  &---    &---   &---     &--- &---  \\ \hline
           &          &      &        &    & \\
$P^{+} P^{-}$  &---    &$400$  $^d$   &$41.7$  $^e$
&$100$  $^f$  &$500$  $^f$  \\ \hline \hline
           &          &      &        &   & \\
$P_8^{0 \prime} (\eta_T)$  &$400 - 500$  $^g$    &$325$  $^h$
  &--- &---     &---   \\ \hline
           &          &      &        &  &  \\
$P_8^{0}$  &$10 - 20$  $^h$     &$325$  $^{h,i}$   &---     &---
                                                    &--- \\ \hline
           &          &      &        &   &  \\
$P^{+}_8 P^{-}_8$   &$10 - 20$  $^{h,i}$    &$325$  $^{h,i}$   &$45$
 $^e$
         &$100$  $^f$  &$500$  $^f$  \\ \hline \hline
           &          &      &        &   &  \\
$P^{+}_3 P^{-}_3$   &---$^i$    &---$^i$   &---
         &$100$  $^f$  &$500$  $^f$  \\ \hline \hline
\end{tabular}
\end{center}
\end{table}
\noindent
$^a$ Decay mode $P^{0 \prime} \rightarrow \gamma \gamma$ ,
similar to a light neutral Higgs \cite{TDR}. \\
$^b$ Decay mode $Z \rightarrow \gamma P^{0 \prime}$,
assuming a one-family model, with $N_{TC} = 7$ and $N_{TC}= 8$ respectively;
no reach for $N_{TC} < 7$; for larger $Z \gamma P^{0 \prime}$ couplings,
the discovery reach extends to $65$ GeV \cite{ehsAMLR,ehsLRES,ehsVL}. \\
$^c$ No reach for traditional one-family model;
possibility of reach for the Lane-Ramana~
\cite{Lane1991} multiscale model in several processes. The
discovery reach could be greatly improved if the TLC  operates
in a $\gamma \gamma$ mode. \\
$^d$ Estimated from work on charged Higgs detection (via
$g b \rightarrow t H^- \rightarrow t \overline{t} b$)
for $\tan \beta \simeq 1$, $m_t = 180$ GeV, $100 \; \mbox{fb}^{-1}$
integrated luminosity and assuming a
$b$-tagging efficiency $\epsilon_b = 0.3$~\cite{hplus}. \\
$^e$ ALEPH and DELPHI limit \cite{ALEPH1992}, while the OPAL
limit \cite{ehsOPAL} is $M(P^\pm) > 35\; GeV$.
The kinematic limit in LEP I is $M_Z/2$.\\
$^f$ Kinematical limits for LEP200 and a $1$ TeV $e^+ e^-$ collider (TLC)
\cite{kin}. \\
$^g$ Contribution to the $\overline{t} t $ cross section in multiscale models
\cite{ehsEL}. \\
$^h$ QCD pair production of colored PGBs with decay into $4$ jets
\cite{ehsCSG}.\\
$^i$ QCD pair production of colored PGBs, each decaying to $t\overline t$,
$t\bar b$, $t\tau$ or $t\nu_\tau$ should allow higher reach in mass.
This has yet to be studied.

%%%%%%%%%%%%%%%%%%%%%%%%%%%
%%%%%%%%%%%%%%%%%%%%%%%%%%%
\section{Oblique corrections and parameters S through X}

\hspace{0.5cm}
Although the investigations about the new physics beyond the SM
has been continued for many years, no any new particles beyond
those predicted by the SM (such as the Z and W gauge bosons and the
top quark) have been discovered by various experiments.
If the new physics is too heavy to be directly produced in current
experiments, there are generally two ways for it to indirectly
contribute. It can  contribute  to:
{\it
\begin{quotation}
(a). the propagation of the gauge bosons ( $\gamma,\; Z^0$ and $W^{\pm}$),
{\em e.g.,} the so-called ``Oblique'' corrections;

(b). the three point fermion-boson  and/or the four point fermion-fermion
interactions, {\em e.g.,} the  ``Non-oblique'' corrections;
\end{quotation}}

In this section we present a brief review for the definitions and
estimations of the six oblique parameters S
through X.

%%%%%%%%%%%%%%%%%%%%%%%%%%%
\subsection{Oblique parameters $(S, T, U, V, W, X)$}

\hspace{0.5cm}
In general, if only the ``oblique'' contributions from new physics are
considered, one can write the self-energy
functions of gauge bosons as a summation of the SM part and the new physics
part:
\begin{eqnarray}
\Pi_{ab}(q^2) = \Pi_{ab}^{SM}(q^2) + \delta \Pi_{ab}(q^2),\ \
with\ \ (a,\;b)=(ZZ,\; WW,\; \gamma \gamma,\; Z\gamma ),
\end{eqnarray}
where the first term represents the SM contributions, while all new-physics
``oblique'' corrections are contained in the second term.

The oblique corrections have been very conveniently
parametrized in terms of three parameters S, T and U (the U parameter is
small and usually can be ignored) by Peskin and Takeuchi \cite{Peskin}.
They assumed that the new particles running round the loops have large
masses ( much
larger than the masses of the W and $Z^0$)  so that
the self-energies could be described well by a Taylor expansion to
linear order: $\delta \Pi_{ab} \approx A_{ab} + B_{ab}q^2$, the errors of
order $(M_Z^2/M_{new}^2)$ were neglected.

Under this
approximation Peskin and Takeuchi \cite{Peskin} estimated S in TC theory from
a scaled-up QCD dispersion relation
and concluded that $S\approx 1.6 $ for the OGTM and
$S\approx 0.5$ for the ODTM (assuming $N_{TC}=4$
in both cases), while the fitting of the data (done at 1991) predicted
$S=-1.52\pm 0.84$ \cite{Peskin}.
But the situation has been changed recently, a new fit (done at 1994)
of the electroweak data
leads to $S=-0.21\pm 0.24 ^{-0.06}_{+0.17}$ \cite{Langacker}, which is
close to zero with small error,
and the tendency to find $ S < 0 $ that existed in earlier data
is no longer present.

Burgess {\it et al}  \cite{Burgess1} extended the $(S,T,U)$
parametrization by introducing three additional parameters $(V,W,X)$
to describe the lowest non-trivial momentum dependence in oblique
diagrams.
If the heavy new physics assumption is dropped, the gauge-boson
self-energies have some complicated dependence on $q^2$ that cannot
be adequately expressed using the first few terms of a Taylor expansion.
Nonetheless, since precision observables are associated
only with the scales
$q^2 \approx 0$, $q^2 = M_Z^2$ or $q^2 = M_W^2$, it turns out that
it is possible
in practice to parametrize oblique effects due to light new physics
in terms of only six parameters $S$,
$T$, $U$, $V$, $W$ and $X$. These are defined as
\cite{Burgess1,Evans1,Burgess2}
\beq
\alpha S &=& -4 s_w c_w (c_w^2-s_w^2)\dwpi_{ZA}(0)
-4 s_w^2 c_w^2 \dwpi_{AA}(0)\nonumber\\
&&\ \ \ \ +  4 s_w^2 c_w^2 \left [ \frac{\dpi_{ZZ}{(M_Z^2)}
-\dpi_{ZZ}{(0)}}{M_Z^2}
\right ], \label{eq:ss}\\
\alpha T &=& \frac{\dpi_{WW}(0)}{M_W^2} -\frac{\dpi_{ZZ}(0)}{M_Z^2}
\label{eq:tt}\\
\alpha U &=& 4 s_w^2 \left [ \frac{\dpi_{WW}{(M_W^2)}-\dpi_{WW}{(0)}}{M_W^2}
\right ]
-4 s_w^2 c_w^2 \left [ \frac{\dpi_{ZZ}{(M_Z^2)}-\dpi_{ZZ}{(0)}}{M_Z^2}
\right ] \nonumber\\
&&\ \ \ \ -4 s_w^4 \dwpi_{AA}(0) -8 c_w s_w^3 \dwpi_{ZA}(0),
\label{eq:uu}\\
\alpha V &=& \dpi_{ZZ}^{'}(M_Z^2) -
\left [ \frac{\dpi_{ZZ}{(M_Z^2)}-
\dpi_{ZZ}{(0)}}{M_Z^2} \right ],\label{eq:vv} \\
\alpha W &=& \dpi_{WW}^{'}(M_W^2) -
\left [ \frac{\dpi_{WW}{(M_W^2)}-
\dpi_{WW}{(0)}}{M_W^2} \right ],\label{eq:ww} \\
\alpha X &=& -s_w c_w
\left [ \dwpi_{ZA}(M_Z^2)- \dwpi_{ZA}(0)\right ], \label{eq:xx}
\eeq
where $\dwpi(q^2)\equiv \dpi(q^2)/q^2$, and where $\dpi'(q^2)$ denotes
the ordinary derivative with respect to $q^2$.
The $V$, $W$ and $X$ are intentionally defined so that they vanish
when the self-energies are linear functions of $q^2$ only,
in which case the $STU$ parametrization is exactly recovered.
For the questions of how the above parameters appear
in expressions for Z-pole observables, the reader can see
the refs.\cite{Burgess1,Evans1,Burgess2,Fleming1}.

A global fit (done at the end of 1993) to the data in which
all six oblique parameters S through X are allowed to vary
simultaneously gives the one standard deviation bounds \cite{Burgess2}:
\beq
S \sim -0.93 \pm 1.7, \ \ \ V \sim 0.47 \pm 1.0, \nonumber\\
T \sim -0.67 \pm 0.92, \ \ \ W \sim 1.2 \pm 7.0, \label{eq:sxexp}\\
U \sim -0.60 \pm 1.1, \ \ \ X \sim 0.1 \pm 0.58. \nonumber
\eeq
{}From eq.(\ref{eq:sxexp}), it is easy to see that the inclusion
of $V$, $W$, and $X$ weakens the bounds on $S$, $T$, and $U$
considerably. This analysis raises the possibility that a TC model
with new light particles with masses of order $M_Z$
may be experimentally viable. There are two possible sources of such light
particles: light technifermions and the light PGBs that occur in many
TC models with large global symmetries. In ref.\cite{Evans1}, N.Evans
estimated the possible contributions to parameters $V$, $W$, and $X$ from
the light technifermions and pseudo-Goldstone bosons. For the OGTM,
the inclusion of new contributions could relax the upper bounds on
$S$ and $T$ by between 0.1 and 1 depending upon the precise particle
spectrum. For more details the reader can see the original
paper \cite{Evans1}.

In ref.\cite{Bamert2}, the authors argue that the oblique corrections to
all Z-pole observables can
be expressed in terms of only two parameters, $S'$ and $T'$, which are linear
combinations of $S$ through $X$:
\beq
S' &=& S + 4 (c_w^2- s_w^2)X + 4 s_w^2 c_w^2 V, \label{eq:sprim}\\
T' &=& T + V\label{eq:tprim}
\eeq
The effective vertex for neutral currents at the Z-pole is now given by
\beq
i\Lambda^\mu_{\rm nc}(M_Z^2)
&=& - i \; \frac{e}{s_w c_w} \; ( 1 + {1 \over 2}\alpha T')\; \gamma^\mu
 \cdot\nonumber\\
& & \ \  \left[ I_3^f \gamma_L - Q^f \left( s_w^2 +
\frac{\alpha S'}{ 4 ( c_w^2 -  s_w^2)} -
\frac{ c_w^2 s_w^2 \; \alpha T'}{c_w^2 - s_w^2}  \right) \right]
\label{eq:effzpole}.
\eeq
So, in confronting
some model of light new physics with $Z$-pole data, one
would calculate $S'$ and $T'$ rather than $S$ and $T$.
With $S'$ and $T'$ defined this way, the low-energy neutral-current
observables
now depend on $S', T', V$, and $X$; the $W$-mass depends on
$S', T', U, V$, and $X$.
Fits to the most recent LEP and SLC data (Winter 1995) are presented in
\cite{Bamert}, the result is
\beq
S'& = & \; - 0.20 \pm 0.20,  \ \ \ T' =  - 0.13 \pm 0.22 \nonumber\\
\alpha_{{\rm s}}(M_Z) &= & \; 0.127 \pm 0.005\label{eq:95fit}
\eeq

%%%%%%%%%%%%%%%%%%%%%%%%%%%%%%%%
\subsection{ Estimations of S through X in the OGTM }

\hspace{0.5cm}
Generally speaking, the contributions to the parameters $S$ through $X$
(in most cases only $S$ was considered)
in the OGTM can be divided into two parts: the `high-energy' part from the
techniquarks and technileptons, and the `low-energy' part from those PGBs.
It is well known that, only a few years ago,
oblique correction considerations hinging on the parameter $S$
tended to rule out certain models of
Technicolor \cite{Peskin,King}.

The $S$-argument against Technicolor was countered in
ref.\cite{Appel}, where it was pointed out that the
high-energy contribution determined
from scaling the parameters of the QCD chiral lagrangian represents an upper
bound, and that other methods used to estimate this contribution result in a
smaller or negative value for the high-energy piece. The authors of
ref.\cite{Appel}
stated that the isospin splitting and
techniquark-technilepton splitting in the OGTM can
reduce the predicted value of the electroweak
parameter $S$, without making a large
contribution to the $T$ parameter. they
naively estimate the high-energy contribution by calculating the one
loop technifermion diagrams, and find that, after adding it
to the low-energy piece, the $S$-argument against Technicolor can be
invalidated. Thus, ref.~\cite{Appel}, entitled
``Revenge of the one-family Technicolor
models," re-established the possible phenomenological
viability of this model.

In ref.\cite{Fleming2}, the authors examined the oblique
correction phenomenology of
one-family technicolor model with light pseudo-Goldstone
bosons.   From loop calculations based on a gauged chiral
lagrangian for Technicolor, they conclude that
even though loops with light Goldstone bosons give
a negative contribution to $S$ measured at the $Z$-pole, this effect is not
sufficiently large to unambiguously
counter the `S-argument' against one-family Technicolor.

Using the effective lagrangian method, the authors \cite{Fleming2}
explicitly calculated the one-loop oblique corrections to electroweak
parameters
form different sources. We here only list the main results presented
in ref.\cite{Fleming2}, for more details the reader can see the original
paper.

$\bullet$\hspace{.5cm} The ``high-energy'' contribution is large and
positive:
\beq
S(\Lambda_{TC}) = \;
- 16 \pi \; {N_d N_{TC}\over N_{QCD} } \;
L_{10}^{QCD}(\Lambda_{QCD}) \sim +1   .
\eeq

$\bullet$ \hspace{.5cm}
The contribution from Isotriplet PGBs is positive in sign,
and its size depends on the details of particle spectrum:
\beq
 \alpha S ( isotriplets) = {e^2 \over 24 \pi^2}
\log{\Lambda^2_{TC} \over M_Z^2} +
 {\hbox{convergent pieces}}\label{eq:stc}
\eeq

$\bullet$ \hspace{.5cm}
According to the calculations carried out in ref.\cite{Fleming2},
the contribution to the $S$ parameter from the non-self-conjugate
isosinglets is generally negative, and
there is no contribution from a self-conjugate isosinglet.
For $m_\pi = M_Z/2$, one has
\beq
\alpha S = \; - \; {e^2 s_w^4 y^2\over 2 \pi^2}\;\left( {1 \over 9} \right),
\label{lightpgbs}
\eeq
and for $m_\pi \gg M_Z$, one  has
\beq
\alpha S = \; - \; {e^2 s_w^4 y^2\over 2 \pi^2}\;
\left( {1 \over 60}\; { M_Z^2\over m_\pi^2} \right) .
\eeq
This negative value could be taken as a reassuring sign
if one wanted to further
establish the phenomenological feasibility of Technicolor.
However, it must be appreciated that
of the 60 physical PGBs in one-family Technicolor,
only three pairs of particles
are non-self-conjugate singlets as illustrated in Table \ref{pgbs}.
The great majority of the
PGBs are
arranged in triplets, and therefore the negative $S$ contributions
from the few non-self-conjugate singlets cannot
effectively counter the positive contributions from the
many triplets. However, there are ways out. If the isotriplets
are heavy as predicted in ref.\cite{Xiao4} their positive
contribution to $S$
should be very small.  While the negative contributions to $S$ from those
light isosinglet PGBs may be large in size as given in eq.(\ref{lightpgbs})
if the isosinglets of PGBs are sufficiently light.
Under these circumstances the negative corrections from
light isotriplets would dominate.

In my opinion whether the ``S argument''
against Technicolor can be avoided or not is still unclear,
and therefore,
further investigations about this problem are still needed.

%%%%%%%%%%%%%%%%%%%%%%
%%%%%%%%%%%%%%%%%%%%%%

\section{Non-oblique corrections on $\zbb$ vertex in TC models}

\hspace{0.5cm}
Now we turn to study the non-oblique corrections to the physical
observables in TC models \cite{Farhi,Weinberg}. Of cause, other new physics
models also can contribute to the observables in different ways,
such as corrections from the
mirror particles in the Minimal Supersymmetric Standard Model(MSSM)
\cite{MSSM} or other beyond models \cite{Beyond},
but we here don't deal with such models.

In the process of ETC gauge group breaking,
many ETC gauge bosons become massive. Some of them called ``sideways''
cause the transition of the ordinary fermions to the technifermions,
some of them called ``horizontal'' connect the ordinary fermions themselves,
and the others called ``diagonal'' diagonally interact with both the
ordinary fermions and technifermions.

There are two kinds of sources of non-oblique corrections to the
$\zbb$ vertex in non minimal TC models, namely from
ETC gauge boson exchange \cite{Simmons,Kitazawa,Wu}
and from charged PGB exchange \cite{Xiao1,Xiao2}, and we will discuss
these two kinds of corrections in the following subsections,
respectively.

%%%%%%%%%%%%%%%%
\subsection{Negative contributions from sideways ETC boson exchange}

In ref.\cite{Simmons} R.S.Chivukula {\it et al}
have estimated the non-oblique effects in the $\zbb$ vertex from
sideways ETC gauge boson exchanges. If the top quark mass is
generated by the exchange of an
$SU(2)_W$ neutral ETC gauge boson (the most popular case for ETC models)
with mass $M_{ETC}$, then this gauge boson carries technicolor and couples
with strength $g_{ETC}$ to the current
\beq
\xi\, \overline{\Psi}_L^i\, \gamma^{\mu}\,T_L^{iw}
+(\frac{1}{\xi})\,\overline{t}_R\, \gamma^{\mu}\,U_R^{w},
\label{eq:etccoupling}
\eeq
where $\overline{\Psi}_L=(t,b)_L$, $T_L=(U,D)_{L}$ with U and
D technifermions,
the indices i and w are for $SU(2)_W$ and technicolor, respectively.
The constant $\xi$ is the {\it Clebsch-Gordon}-like coefficient of order
one associated with the ETC gauge group. The top mass is then given by
\beq
m_t=\frac{g_{ETC}^2}{M_{ETC}^2}\;<\overline{U}U>
\approx \frac{g_{ETC}^2}{M_{ETC}^2}\cdot 4\pi F_{\pi}^2
\label{eq:mt}
\eeq
where the condensate, $<\overline{U}U> $, has been estimated by
naive dimension
arguments \cite{Manohar} in terms of the technipion decay constant,
$F_{\pi}=246/\sqrt{N_D}$ with  $N_D$ is the number of technifermion doublets.

As described in ref.\cite{Simmons} the ETC interactions in
eq.(\ref{eq:etccoupling})
can give rise to a correction
\beq
\delta g_L=-\frac{\xi^2}{2} \frac{g_{ETC}^2 F_{\pi}^2}{M_{ETC}^2}
\frac{e I_3}{s_w c_w}\label{eq:dgl1}
\eeq
to the tree-level $\zbb$ coupling $g_L$. Substituting for
$g^2_{ETC}/M_{ETC}^2$ form eq.(\ref{eq:mt}) one finds
\beq
\delta g_L^{ETC} \sim \frac{1}{4}\cdot \frac{m_t}{4\pi F_{\pi}}\cdot
\frac{e}{s_w c_w}.\label{eq:dgl2}
\eeq
This correction $\delta g_L$ can result
in a contribution to the  $\zbb$ vertex,
as given in ref.\cite{Simmons},
\beq
\Delta_b^{ETC}\;(sideways) \approx -6.6\%\times \xi^2\cdot \left [
\frac{m_t}{180GeV} \right ].\label{eq:deltetcs1}
\eeq

For the ODTM, no Pseudo-Goldstone bosons can be survived when the
chiral symmetry was broken by the condensate $<T\overline{T}> \neq 0$,
but the sideways ETC gauge boson exchange can produce typically large and
negative contribution as illustrated in eq.(\ref{eq:deltetcs1}).
Although the ODTM is only a toy
model in nature the correction in eq.(\ref{eq:deltetcs1})
is universal for most popular TC/ETC models with standard ETC
dynamics($e.g.$ the ETC gauge boson is $SU(2)_W$ singlet).

%%%%%%%%%%%%%%%%%%%%%%%%%%

\subsection{Positive contributions from diagonal ETC boson exchange}

\hspace{0.5cm}
The sideways ETC gauge bosons must exist
in the realistic model to generate the quark and lepton masses,
while the existence of diagonal ETC gauge bosons   is model-dependent.
Lightest ETC bosons are the sideways and diagonal ETC gauge bosons
associated with the top quark.

In ref.\cite{Kitazawa} Kitazawa calculated the radiative
contributions to the $\zbb$ vertex generated by the diagonal ETC
gauge boson exchange. He found that the diagonal ETC gauge boson
also yields non-oblique correction through the mixing with Z boson,
and both kinds of contributions (sideways and diagonal) are {\em positive}
and don't cancel each other. The diagonal contribution is $30\%$ of the
sideways contribution when $\xi_t=1$.

Very recently, Wu \cite{Wu} reconsidered the non-oblique corrections
on the $\zbb$ vertex from diagonal ETC gauge boson exchange. He found
that the diagonal ETC gauge boson
exchange really contribute to the $\zbb$ vertex as calculated by Kitazawa,
but the contributions from the sideways and diagonal ETC gauge boson
exchanges are { \em opposite in sign }, and therefore, {\em these two kinds
of contributions will cancel each other}.

According to the calculations in ref.\cite{Wu},
for one-generation TC model the diagonal ETC gauge
boson exchange could  result in a  contribution the tree-level
$\zbb$ coupling \cite{Wu}:
\beq
\delta g_L^b(diagonal) \approx -\frac{f^2_Q}{8 m^2_{X_D}}
\frac{N_C}{N_{TC}+1} \;g_{E,L}\;(g^U_{E,R} - g^D_{E,R}),
\label{eq:deltaglb2}
\eeq
where the $N_C=3$ is the number of colors, the $N_{TC}$ is the number of
technicolors, and the definitions of all other parameters appeared in above
equation can be found in ref.\cite{Wu}. The result of eq.(\ref{eq:deltaglb2})
differs by a minus sign from the loop estimate of ref.\cite{Chivukula2}.
Summing up the sideways and diagonal ETC exchange contributions gives:
\beq
\delta g_{L,ETC}^b \approx  -\frac{f^2_Q}{8} \left [
\frac{g_{E,L}^U\,(g^U_{E,L}-g^D_{E,L})}{m^2_{X_D}}
\frac{N_C}{N_{TC}+1}-\frac{g_{E,L}^2}{m^2_{X_S}} \right ].
\eeq
It is seen from the above expression that the two contributions are of
comparable magnitude but with opposite sign, and they will basically
be canceled out for proper choice of parameters. It is also possible
for ETC exchange to give a small positive correction to $R_b$.

Obviously, because of the cancellation of these two kinds
of contributions, the ETC-corrected $R_b$ value could lie in a range
consistent with the LEP data {\it if there were no other kinds of
corrections on this ratio}.
Although there are some differences between the one-generation TC model
studied in ref.\cite{Wu} and the QCD-like OGTM, the basic structures
(such as
the gauge group of the model, the particle spectrum, the couplings,
$\cdots$, etc) are
very similar. And therefore we can assume that the diagonal ETC gauge boson
exchange in the ordinary OGTM may
produce the similar positive contributions to the $R_b$, at least the total
ETC non-oblique correction is very small.

We know that, however, besides the ``high energy''
contributions to the $\zbb$ vertex from ETC gauge boson exchanges,
there are also ``low-energy'' negative
contributions from the charged PGBs as estimated
in ref.\cite{Xiao1,Xiao2}. These contributions will decrease
the ratio $R_b$ by as large as
a few percent \cite{Xiao1}, and the exact size of corrections from
charged PGBs depend on the top quark mass and the masses of charge
PGBs ( color singlets and color octets).
In spite of some uncertainties in the evaluation of ref.\cite{Wu},
Wu's work is great help for one to extract the lower-limit on the charged
PGBs from the present data because one can now reasonably assume that
the total corrections on the ratio $R_b$ from ETC dynamics are very
small and can therefore be
neglected at first approximation, $e.g.,$ one can assume that:
\beq
\Delta_b^{new}(OGTM) &=&  \Delta_b^{ETC}(sideways) +
\Delta_b^{ETC}(diagonal) +
\Delta^{P^{\pm}}_{bv} + \Delta^{P^{\pm}_8}_{bv} \nonumber\\
&\approx & \Delta^{P^{\pm}}_{bv} + \Delta^{P^{\pm}_8}_{bv}.
\eeq

%%%%%%%%%%%%%%%%%%%%%%%%%%%%%%%%%%%%%
\subsection{Negative contributions from charged PGBs}

\hspace{0.5cm}
In contrast to the ODTM (where there is no PGBs),
the charged PGBs appeared in the OGTM also contribute a negative correction
to the $\zbb$ vertex as estimated in refs.\cite{Xiao1,Xiao2}.
In short, there are three kinds of non-oblique corrections on the
$\zbb$ vertex in the OGTM:
{\it
\begin{quotation}
(a). $\Delta_b^{(top)}$, the correction on the $\zbb$ vertex
arising from loop diagrams
involving the internal heavy top quark, which is the same as in the
standard model;

(b). $\Delta_b^{ETC}$, the correction on the $\zbb$ vertex
from sideways and diagonal ETC gauge boson exchange in the OGTM,
the total corrections is small and can be neglected at first
approximation;

(c). $\Delta_b (PGBs) =\Delta_{bv}^{P^\pm} + \Delta_{bv}^{P_8^\pm}$,
the corrections on $\zbb$ from the color singlet and color octet
charged PGBs.
\end{quotation}}

In ref.\cite{Xiao1,Xiao2} we calculated the non-oblique corrections to
the $\zbb$ vertex from the charged PGBs and obtained the lower limits on the
color octet PGBs. We here just discuss this work briefly.

The gauge couplings of the PGBs to the gauge bosons
$(\gamma,\; Z, \; W^\pm)$ are determined by their quantum numbers.
The coupling of PGBs to ordinary fermions are induced by ETC
interactions and hence are model dependent. However, these couplings
are generally proportional to the fermion masses.
In ref.\cite{Ellis1981} J.Ellis {\it et al} estimated the Yukawa couplings
to ordinary fermions of the PGBs in the OGTM under some simplifying
assumptions. In their first Monophagic Model,
the ETC generators commute with $SU(3)_C \otimes SU(2)_L \otimes U(1)_Y$
and couple each type of ordinary fermions to the same type
of technifermions in the sense of avoiding the FCNC problem.
The effective  Yukawa couplings of the charged color singlet(color octets)
PGBs $P^{\pm}$( $P^{\pm}_8$) are the form of
\beq
\left (\frac{-i}{F_\pi}\right )\, P^+\, \left [\,
\overline{u}\,(V_{km}\, m^d \,
\frac{1+\gamma_5}{2}
-m^u\, V_{km} \, \frac{1-\gamma_5}{2})\, d \, \sqrt{\frac{2}{3}}\right]
+ H.C.
\label{eq:coupling1}\\
\left (\frac{-i}{F_\pi} \right )\, P^+_{8\alpha}\,
\left [\,\overline{u}\,(V_{km}\, m^d\:
\frac{1+\gamma_5}{2}
-m^u \: V_{km}\, \frac{1-\gamma_5}{2})\, \lambda^\alpha\, d\, \right ] 2
+ H.C. \label{eq:coupling2}
\eeq
where the $V_{km} $ is the element of KM matrix. In these two effective
couplings the Goldstone boson decay constant $F_{\pi}$ is
$F_{\pi}= 246/ \sqrt{N_D} = 123\, GeV$
in order to ensure the correct masses for the gauge bosons
$Z^0$ and $W^{\pm}$.

Based on the effective Yukawa couplings as shown in
eqs.(\ref{eq:coupling1}, \ref{eq:coupling2}) and  the $ZP^+P^-$
couplings given in ref.\cite{Chadha1981}
we can write down the Feynman rules
 needed in the calculation for the TC correction
to $Zb\overline{b}$ vertex at one-loop order.
\beq
{\bf [Z-b-\overline{b}]}\ \ &=& ie\gamma^{\mu}(v_b-a_b\gamma_5) \\
{\bf [P^+-t-b]}\ \ &=&i\frac{V_{tb}}{2F_\pi}\cdot \sqrt{\frac{2}{3}}
[m_t(1-\gamma_5)- m_b(1+\gamma_5)] \\
{\bf [P^+_8-t-b]}\ \ &=& i\frac{V_{tb}}{2 F_\pi}\cdot 2\lambda^{\alpha}
[m_t(1-\gamma_5)- m_b(1+\gamma_5)] \\
{\bf [Z-P^+-P^-]}\ \ &=& ie\frac{1-2s_w^2}{2s_wc_w}(P^+ - P^-)^{\mu} \\
{\bf [Z-P^{+}_8-P^-_8]}\ \ &=& ie\frac{1-2s_w^2}{2s_wc_w}
(P^+ - P^-)^{\mu}\cdot \delta_{\alpha \beta}
\eeq
where the $\lambda^\alpha$ are the Gell-Mann $SU(3)_c$ matrices and the
vector and axial vector coupling constants for bottom quark  are:
\beq
v_b=\frac{-\frac{1}{2}+\frac{2}{3}s_w^2}{2s_w c_w},
\ \ a_b= \frac{-1}{4s_w c_w}
\eeq

After the analytical calculation of those relevant Feynman diagrams as
shown in Fig.1 of ref.\cite{Xiao1} we got an effective $\zbb$ vertex:
\beq
&& ie\gamma^{\mu}(v_b - a_b\gamma_5)
+ ie\gamma^{\mu}(1 - \gamma_5)\frac{A^2}{16\pi^2}\cdot |V_{tb}|^2\nonumber\\
&&\ \ \ \ \ \cdot \{ \; \left [\, F_{1a} + F_{1b} + F_{1c}\,\right]
+ \left[ F_{8a} + F_{8b} + F_{8c}\right ] \cdot 6\lambda^{\alpha}
\lambda^{\alpha} \; \} \label{eq:zffpgbs}.
\eeq
The explicit expressions of the form factors $F_{1a}, F_{1b},F_{1c}$
and $F_{8a}, F_{8b}$,$F_{8c} $ can be found in ref.\cite{Xiao1}.

Using the effective $Zb\overline{b} $ vertex as given in
eq.(\ref{eq:zffpgbs}),
it is straightforward to calculate the contributions to the partial width
$\Gamma_b$ and $R_b$ from the charged PGBs. We get the results:
\beq
\delta \Gamma_{TC} = \Gamma^{(0)}_b [\Delta_{bv}^{P^{\pm}} +
\Delta ^{P^{\pm}_{8}}_{bv}]
\eeq
where $\Gamma^{(0)}_b=380 \; MeV$,  and the vertex
factors are the form of
\beq
\Delta^{P^{\pm}}_{bv} &=& \frac{A^2s_wc_w}{4\pi^2}\,|V_{tb}|^2
\frac{(3+\beta^2)(v-1)+ 3(1-\beta^2)(v+1)}{(3-\beta^2)v^2 + 2\beta^2}
\nonumber\\
&&\ \ \ \cdot \left [\,ReF_{1a}+ReF_{1b} + ReF_{1c}\,\right ],
\label{eq:deltabva}\\
\Delta^{P^{\pm}_8}_{bv} &=& \frac{6\,A^2s_wc_w}{4\pi^2}\,|V_{tb}|^2\,
T(8)\cdot \frac{(3+\beta^2)(v-1)+ 3(1-\beta^2)(v+1)}{(3-\beta^2)v^2
+ 2\beta^2}\nonumber\\
&& \ \ \ \cdot \left [\,ReF_{8a}+ReF_{8b} + ReF_{8c}\,\right ]
\label{eq:deltabvb}
\eeq
where the $s_w$ and $c_w$ are the mixing angle, and
the explicit expressions of all other parameters in
eqs.(\ref{eq:deltabva}, \ref{eq:deltabvb})
can be found in ref.\cite{Xiao1}.

%%%%%%%%%%%%%%%%%%%%%%%%%%

\subsection{Updated constraints on masses of PGBs}

\hspace{0.5cm}
As described in ref.\cite{Xiao1}, the magnitude of the  vertex factors
$\Delta_{bv}^{P^\pm}$ and
$\Delta_{bv}^{P_8^\pm}$ depends on three parameters: the top quark mass
$m_t$, the mass of color-singlet PGBs $m_{p1}$ and the mass of color-octet
PGBs $m_{p2}$. Therefore, the exact size of this kind of corrections depend
on the  mass spectrum of top quark and charged PGBs.
Historically, theoretical estimations about the masses of charged PGBs
have been done by many authors\cite{Peskin2,Farhi}.
We here  discuss briefly the constraints on
the masses of color-singlet
charged PGBs $P^{\pm}$ and color-octet charged PGBs $P_8^{\pm}$ because
only these two kinds of PGBs can contribute significantly
to the $\zbb$ vertex.

Using the SIP, it is straightforward to calculate the values of
$\Delta_{bv}^{P^\pm}$ and  $\Delta_{bv}^{P_8^\pm}$ from
eqs.(\ref{eq:deltabva}, \ref{eq:deltabvb}).
For $m_t=180\;GeV$,
\beq
\Delta_{bv}^{P^\pm} &=& (-0.013 \sim -0.002),\; \hbox{for
$m_{p1}=50 - 400 \;GeV$},  \label{eq:dbvna} \\
\Delta_{bv}^{P_8^\pm} &=& (-0.050 \sim -0.003),\; \hbox{for
$m_{p2}=200 - 650 \;GeV$}.\label{eq:dbvnb}
\eeq
The contributions from the charged PGBs are always negative and will push
the OGTM prediction for the vertex factor $\Delta_b^{new}$ away from the
measured $\Delta_{b,exp}^{new}$.
These negative corrections are clearly disfavored by the current
data. But fortunately, the charged PGBs show a clear decoupling
behavior as listed in eqs.(\ref{eq:dbvna}, \ref{eq:dbvnb}).

In the OGTM, the total size of vertex factor $\Delta_b$ generally depend on
two " free" parameters, the masses $m_{p1}$
and $m_{p2}$ if we use $m_t=180\pm 12$ GeV as input.
The current data
will enable us to exclude large part of parameter space
of $m_{p1}$ and  $m_{p2}$ in the $m_{p1}-m_{p2}$ plan, as shown in Fig.2.
{}From Fig.2 one can read out the bounds on the masses $m_{p1}$ and $m_{p2}$,
\beq
m_{p1} > 200\;GeV, \ \ \hbox{for ``free''}\ \ m_{p2},\label{eq:limita}
\eeq
and
\beq
m_{p2} > 600\;GeV, \ \ \hbox{ for  $ m_{p1} \leq 400\;GeV$}.\label{eq:limitb}
\eeq
while the uncertainties of $m_t$, $\delta m_t=12\;GeV$, almost don't
affect the constraints.  The lower limit on $m_{p1}$ as given in
eq.(\ref{eq:limita}) is the highest lower limit derived so far from
the precision data. The lower limit on $m_{p2}$ in eq.{\ref{eq:limitb})
is much stronger than that has been given before in ref.\cite{Xiao2}.
The inclusion of the remained corrections from ETC dynamics in the OGTM will
alter (strengthen or waken ) the bounds on $m_{p1}$ and $m_{p2}$, but this
ETC effect will be small and model-dependent.

According to our studies we can conclude
that the charged PGBs should be much heavier than that estimated before
and these heavy charged PGBs most probably decouple from the ``low-energy''
(e.g.,
the $M_Z$ scale) physics, if they were existed indeed.

Of cause, the lower limits on charged PGBs depend on the effective
couplings of the model being studied. For non-QCD-like TC models, the
effective couplings may be different from those as given
in ref.\cite{Ellis1981,Chadha1981}, and consequently the lower limits may
be changed for those models. But I think, the lower limits given here
at least can be viewed as a naive estimation for the masses of
Charged PGBs appeared even in more realistic TC models.

%%%%%%%%%%%%%%%%%%%%%%%%%
\subsection{Non-Oblique Corrections on some processes from PGBs}

\hspace{0.5cm}
Except the non-oblique corrections on the $\zbb$ vertex as discussed
in previous subsections, the PGBs in TC models also contribute to other
physical processes, such as the top quark rare decay, the high-energy
neutral current productions of $t \bar t$ of $b \bar b$ pairs, $\cdots$, etc.
In this section we will give a brief review about the relevant works.

\vspace{.2cm}
$\bullet$\hspace{.5cm} {\bf Top quark rare decays}

 The top quark rare decay $t\rightarrow cV$ has been studied by several
groups in different theories \cite{Wang1994,Eilam1991,Li1994}.
In ref.\cite{Wang1994}, we calculated the vertex corrections to the top quark
rare decays, such as $t\rightarrow cV$ and $t\rightarrow c P^0$ ( where the
V represents the photon $\gamma$, QCD gluons $g$, and gauge boson $Z^0$),
from the PGBs appeared in the OGTM.
We found that these new contributions from the PGBs could enhance the SM
branching ratios by as much as $3 \sim 4$ orders of magnitude for the
favorable parameter space, as illustrated
in Table \ref{Wangtable}. For more details please
see the original paper \cite{Wang1994}.

\begin{table}[htbp]
\begin{center}
\caption{The maximum branching ratios of top quark rare decays
predicted by different models}
\label{Wangtable}
\vskip.5pc
\begin{tabular}{|c|c|c|c|c|c|}   \hline
  & SM $^{\cite{Eilam1991}}$ & 2HDM $^{\cite{Eilam1991}}$
&  QCD$^{\cite{Li1994}}$
& Charginos$^{\cite{Li1994}}$ & PGBs$^{\cite{Wang1994}}$ \\ \hline
$Br(t\rightarrow cZ)$ & $10^{-12}$ &  $10^{-9}$ & $10^{-9}$ &
$10^{-8}$ & $10^{-7}$ \\ \hline
$Br(t\rightarrow c \gamma)$ & $10^{-12}$ &  $10^{-8}$ & $10^{-8}$ &
$10^{-8}$ & $10^{-8}$ \\ \hline
$Br(t\rightarrow cZ)$ & $10^{-10}$ &  $10^{-6}$ & $10^{-6}$ &
$10^{-7}$ & $10^{-6}$ \\ \hline
$Br(t\rightarrow cP^0)$ &  &   &  &
& $10^{-9}$ \\ \hline
\end{tabular}
\end{center}
\end{table}

\vspace{.2cm}
$\bullet$\hspace{.5cm} {\bf The rare decays of $\zbs$}

One of the most characteristic predictions of the SM is
the very small magnitude of FCNC  processes.
Consequently, decays induced by FCNC are an effective way to test the SM,
and, in particular, provide a potentially very sensitive probe of physics
beyond the SM.
Experimentally, the $e^+e^-$ machines can be used as Z
factories providing an opportunity to examine the decay properties of the
neutral weak gauge boson in detail.
The current experimental limit on the FCNC rare Z-decay is about
$10^{-5}$ \cite{Azemoon1992}.
The dominant mode of the flavor changing
Z-decays is
$\zbs$,  and such rare decay
has been thoroughly studied in the SM \cite{Clements1983},
two-Higgs doublet
model \cite{Mukho1990} and in the MSSM \cite{Mukho1989}.
The studies show, in most case, the contributions of these models to
$Br(Z\rightarrow b\overline{s}+\overline{b}s)$ are at the order of $10^{-7}$.
Such contributions are too small to be detected in near future.

In Ref.\cite{Ellis1981}, two kinds of
the OGTM have been studied by Ellis {\em et al}: (a). Model I, the ETC
generators commute
with $SU_L(2)\otimes U_Y(1)\otimes SU(3)_c$ and couple each ordinary fermion
to the technifermion of the same type; (b). Model II,
the ETC/TC generators have
$SU(2)_L$ isospin=0 but couple u,c,t...to $\overline{E}$, d,s,b...e,$\mu,
\tau...$ to $\overline{U}$ and $\nu_e,\nu_{\mu},\nu_{\tau}...$ to
$\overline{D}$).
The flavor-changing decay $\zbs$ induced through PGBs is calculated in
these two kinds of models in ref.\cite{Wang1995}.

Generally speaking, the corrections from the color octet PGBs is
much larger than those coming from the color singlet PGBs, so that
the later one can be neglected.

For  Model I,  we found that an interesting branching ratio
$B(Z\rightarrow b\overline{s}+\overline{b}s)\sim 10^{-6}$ can be obtained
for particular choices of the parameters, such magnitude of the order of
branching ratio is at the border of being detectable.

For Model II,
the PGBS contributions can strongly enhance
the branching ratio $Br(Z\rightarrow b\overline{s}+\overline{b}s)$.
With the current experimental limit on the branching ratios of
rare Z-decay, the constraints on the mass
of color octet Pseudo-Goldstone-bosons can be derived:
$m_p>306$ $GeV$ for $m_t=170$ $GeV$ and $m_p > 333$
$GeV$ for $m_t=190$ $GeV$ respectively. Consequently, the decay $\zbs$
may provide a unique  window to study the TC theory.

\vspace{.2cm}
$\bullet$\hspace{.5cm} {\bf Production of top pairs at NLC}

\hspace{0.5cm}
At next generation $e^{+}e^{-}$ collider (NLC),  operating at
the center-of-mass of $500\; GeV$ with a luminosity of order
$10^{33} cm^{-2} sec^{-1}$ \cite{Peskin1992},
the top quark is dominantly
produced via the process $e^{+}e^{-}\to t\overline{t}$. The radiative
corrections to this process have been calculated in the SM,
in the 2HDM \cite{Been1992} and in the MSSM \cite{Chang1995}.
In ref.\cite{Lu1995} we calculated the TC $0(\alpha m_t^2/m_W^2)$
corrections to this process,
which coming from the virtual effects of PGBs.

Taking into account the TC ${\it O}(\alpha m^{2}_{t}/m^{2}_{W})$
corrections,  renormalized amplitude for
$e^{+}e^{-}\rightarrow t\overline {t}$ is given by:
\beq
M_{ren}\, =\, M_{0}+\delta M^{(\gamma)}+\delta M^{(Z)}
\eeq
where $M_{0}$ is the amplitude at tree level,  $\delta M^{(\gamma, Z)}$
represents the TC ${\it O}(\alpha m^{2}_{t}/m^{2}_{W})$ corrections coming
from the effective $\gamma t\overline{t}$ (Z t$\overline{t}$) vertex.

After the analytical calculations for those relevant Feynman diagrams,
the renormalized
vertices can be expressed in terms of form factors,  which only depend on
the center-of-mass energy $S$ and some masses.
By the numerical calculations we found that the corrections from the
color octet PGBs are dominate.
The one-loop ${\it O}(\alpha m^{2}_{t}/m^{2}_{W})$
 corrections to the physical observables
$\sigma$,  $A_{LR}$ and to the $A_{FB}$
from color octet PGBs can be rather large for relatively  light PGBs.
For $m_t=174\;GeV$ and $m_P=246\;GeV$,
the maximum correction to the observables $\sigma$,  $A_{LR}$ and $A_{FB}$
 can reach $-12.3\%$,  $-11.8\%$ and $-3.3\%$ respectively.
For heavier color octet
PGBs,  the corrections will decrease rapidly(showing a good decoupling
 behavior). Generally speaking,  the corrections to
$\sigma$,  $A_{LR}$ and $A_{FB}$ from Technicolor are  relatively
larger than the others as presented in ref.\cite{Been1992,Chang1995}
and might
be observed at NLC,  since all the  production and decay
form factors of the top quark might be measured at the level
of a few percent at NLC \cite{Peskin1992}.
If any large new physics signals are
received,  these virtual effects of colored-PGBs might provide an possible
 interpretation.

\vspace{.2cm}
$\bullet$\hspace{.5cm} {\bf Production of bottom pairs above Z-pole}

\hspace{0.5cm}
In ref.\cite{Lu1995b}, we
calculated the TC ${\it O}(\alpha m_{t}^{2}/m_{W}
^{2})$ corrections to the process
$e^{+}e^{-} \rightarrow b \overline{b}$ at high
energy $e^{+}e^{-}$ collider above the Z pole. We found that the
corrections from the color octet PGBs dominate.
These TC corrections will affect the size of total cross-section
$\sigma(e^+ e^- \rightarrow b\bar b)$, as well as the
forward-backward asymmetry $A_{FB}$ and left-right asymmetry $A_{LR}$,
\beq
\sigma^{TC} = \sigma_{0}+\delta \sigma^{TC}, \ \ \
\delta A_{LR} = A^{TC}_{LR}-A^{0}_{LR},  \ \ \
\delta A_{FB} = A^{TC}_{FB}-A^{0}_{FB}
\eeq
where $\sigma_{0}$, $A^{0}_{LR}$ and $A^{0}_{FB}$ stand for the values in
the Born approximation. While $\sigma^{TC}$, $A^{TC}_{LR}$ and $A^{TC}_{FB}$
refer to the values with
TC ${\it O}(\alpha m^{2}_{t}/m^{2}_{W})$ corrections.

For the center-of-mass energy $\sqrt{s}$=500 GeV,
$m (P^\pm)=60\; GeV$, $m (P_8^\pm)=200\sim 500$ GeV and $m_{t}=
175 \pm 10\; GeV$, the numerical values of the TC corrections
on the observables
$\sigma$, $A_{FB}$ and $A_{LR}$ from the charged PGBs are the following:
\beq
\frac{\delta \sigma^{TC}}{\sigma^{0}}&=&[\,(-1.9\% \sim -1.0\%),
(-2.6\% \sim -1.3\%),  (-4.2\% \sim -2.6\%)\,]\\
\frac{\delta A_{LR}}{A^{0}_{LR}}&=&[\,(-2.3\% \sim 1.0\%),
(-2.2\% \sim 1.3\%),  (-1.8\% \sim 1.5\%)\,]\\
\frac{\delta A_{FB}}{A^{0}_{FB}}&=&[\,(-1.4\% \sim 1.0\%),
(-1.1\% \sim 1.3\%),  (-0.4\% \sim 1.8\%)\,]
\eeq
{}From the above  results, one can see that the TC correction
on the total cross-section $\sigma(e^+ e^- \rightarrow b\bar b)$
is relatively large for $m_t \approx 180\;GeV$, this effects may
be detectable if the precision of future experiments at NLC
could reach $1\%$ level.
On the other hand,
the corrections  to the
left-right and forward-backward asymmetries are rather small
($\leq 1\%$) if charged PGBs are heavy as shown in eqs.(\ref{eq:limita},
\ref{eq:limitb}).

\vspace{.2cm}
$\bullet$\hspace{.5cm} {\bf Corrections on $BR(B \to X_s \gamma)$
from Charged PGBs}

\hspace{0.5cm}
Recently the CLEO collaboration has observed \cite{Cleo1993} the
exclusive radiative decay $B \rightarrow K^* \gamma $.
The newest upper and lower limits on the branching ratio of $B\to
X_s \gamma$
published by CLEO \cite{Cleo1994} are
\beq
1.0\times 10^{-4} < BR(B \to X_s \gamma) < 4.2\times 10^{-4},
\ \ at \ \ 95\%C.L
\eeq
respectively. As a loop-induced flavor changing neutral current(FCNC)
process the inclusive  decay(at quark level) $b \to s \gamma$ is
in particular sensitive to contributions
from those new physics beyond the Standard Model(SM) \cite{Hewett1994}.

The decay $b\rightarrow s\gamma$ has been investigated within the
framework of Extended
Technicolor(ETC) models by L.Randall and R.S.Sundrum \cite{Randall1993}.
They concluded that the contributions from the ETC gauge boson exchange are
rather small.

In ref.\cite{X1995}, we estimated the possible contributions to the decay
$b\rightarrow s\gamma$
from the exchanges of the charged PGBs $P^\pm$ and $P_8^{\pm}$
with an ordinary
ETC sector. We find that: the new contribution is negative in sign
and the total contribution depends on the values of the
masses of the top quark and those charged PGBs.

If we take experimental result $BR(B \to
X_c e\overline{\nu} ) =10.8\%$ \cite{pdg}, the branching ratios of
$B \to X_s \gamma$ is found to be:
\beq
BR(B \rightarrow X_s \gamma)
\simeq 10.8\%\times \frac{6 \alpha_{QED}\;|C_7^{eff}(m_b)|^2}
{\pi g (m_c/m_b)}
 \left(1-\frac{2 \alpha_{s}(m_b)}{3 \pi} f(m_c/m_b)
\right)^{-1}.
\eeq
where the explicit form of the coefficient $C_7^{eff}(m_b)$
can be found in the original paper \cite{X1995}.
The current $CLEO$ experimental results
can eliminate large part of the parameter space in the
$m(P^\pm) - m(P_8^\pm)$ plan, and specifically, one can put a
strong lower bound
on the masses of color octet charged PGBs $P_8^\pm$:
$m(P^{\pm}_8) > 400\;GeV$ at $95\%C.L$ for free $m(P^{\pm})$.
After we completed this work \cite{X1995} ref.\cite{Balaji1995}
came to our attention, the author also estimated the Technipion
contributions to the rare decay $b \to s \gamma$.

%%%%%%%%%%%%%%%%%%%%%%%%%%
%%%%%%%%%%%%%%%%%%%%%%%%%%
\section{Summary and Conclusions}

\hspace{0.5cm}
In this report we presented a systematic investigation
about the non-oblique corrections on the
$Zb\overline{b}$ vertex from the new physics, such as the
ETC dynamics and the pseudo-Goldstone bosons.
We also discussed the non-oblique corrections
on other processes from the PGBs. According the existed studies one can
expect that the charged PGBs should much heavier than that estimated ever
before.

In my opinion,
Technicolor plus extended technicolor is the most ambitious attempt yet to
explain the physics of electroweak and flavor symmetry breaking and to do
so in natural, dynamical terms.
Of cause, TC and ETC theory also encountered many
problems as discussed in detail in refs.\cite{King,Lane}. It is a difficult
task for TC/ETC theory to explain the large top quark mass and
at the same time
satisfy the constraints from the precision electroweak measurements, such as
the limits from the parameters S and $\Delta_b^{new}$.
But these difficulties do not dissuade me and others from the TC/ETC
philosophy that the origin of this physics is to be found at energies
far below the Planck scale.

The precision data provided by LEP, Tevatron and other high energy colliders
now examine the SM at the loop level. And the accuracy achieved recently
permit us to put some constraints on the existed TC and ETC models,
to pin down the parameter space. On the other hand, the great progress
in the experiments
also encourage the researchers to construct new models with special features,
just like the wind blowing through the calm lake and enforce the water
surface waving!

In recent years, many new TC/ETC models have been constructed
in the sense of avoiding the experimental constraints imposed by the
precision electroweak data. We here simply list ten examples:

\vspace{.2cm}
$\bullet$ \hspace{.4cm}
In ref.\cite{Chivukula}, the authors have shown that a slowly
running technicolor
coupling will affect the size of non-oblique corrections to the
$\zbb$ vertex from ETC dynamics. Numerically, the ``Walking TC''
\cite{Holdom} reduces the magnitude of the corrections at about $20\%$
level. Although this
decrease is helpful to reduce the discrepancy between the TC models and the
current precision  data, however, this improvement is not large enough
to resolve this problem.

\vspace{.2cm}
$\bullet$ \hspace{.4cm}
More recently, Evans\cite{Evans2} points out that
the constraints from $\zbb$ vertex may be avoided if the ETC
scale $M_{ETC}$ can be boosted by strong ETC effects.

\vspace{.2cm}
$\bullet$ \hspace{.4cm}
In ``Non-commuting''
theories ( i.e., in which the ETC gauge boson which generates the top quark
mass does carry weak SU(2) charge), as noted in
refs.\cite{Simmons,Chivukula2},
the contributions on the $\zbb$ vertex come from the
physics of top-quark mass generation and from weak gauge boson mixing
(the signs of the two effects are opposite)\cite{Chivukula2},
and therefore both the size and the sign of the corrections
are model dependent and the overall effect may be small and may even
increase the $\zbb$ branching ratio.

\vspace{.2cm}
$\bullet$ \hspace{.4cm}
In ``TopC assisted TC'' models\cite{Hill1},
potentially low energy top color interactions produce a
top-condensate and accommendate a heavy top quark, while technicolor
is responsible for producing the W and Z masses.
For different options the final result is also different.
As illustrated in ref.\cite{Hill1} the TopC schemes can contain
significant enhancements of the ratio $R_b$, where both the topgluon and the
Z' will provide a positive contribution.
Chivukula {\it et al} very recently discussed some problems
of this model \cite{Chivukula95b}.

\vspace{.2cm}
$\bullet$ \hspace{.4cm}
 ``Low-scale technicolor'',  proposed by King \cite{King1993} with a
low TC confinement scale $\Lambda_{TC} \sim 50 - 100\; GeV$.
Such a low TC scale may
give rise to the first hints of technicolor being seen at LEP I and
spectacular TC signals at LEP 200 and the Tevatron.

\vspace{.2cm}
$\bullet$ \hspace{.4cm}
 ``Technicolor model with a scaler'', constructed by Carone
\cite{Carone1994} and his
collaborators. Although the presence of fundamental scalars seems a retreat
from the original motivation of TC, these kinds of models are worth of
further investigating.

\vspace{.2cm}
$\bullet$ \hspace{.4cm}
 ``Chiral technicolor'', constructed by Terning
\cite{Terning1995}. In this model the technicolor is not vector-like,
but a strongly interacting chiral gauge force. The author proposed a toy
model to demonstrate his new ideas. On the positive side, chiral TC models
offer a simple way to split the $t$ and $b$ quarks without fine-turning.
The ETC contribution to $\Gamma_b$ can be reduced by (up to) a factor of 4,
and the techniqurak contribution to the $S$ parameter can also be reduced.

\vspace{.2cm}
$\bullet$ \hspace{.4cm}
``Realistic one-family TC model''\cite{Appelquist1994},
proposed by Appelquist and Terning. This is an interesting multiscale
Technicolor  model. The reader can see the original paper
\cite{Appelquist1994}.

\vspace{.2cm}
$\bullet$ \hspace{.4cm}
In a new paper \cite{Appelquist1995}, the authors make a connection between
the ETC and the CP-violation problem.
The electric dipole moments of the neutron and the electron in technicolor
theories are estimated to be as large as $\sim 10^{-26}\;e\; cm $
and $\sim 10^{-29}\; e \; cm$, respectively. They also suggest the
potential to observe large CP-violating TC effects in the decay
$t \to W^+ \;b$.  This is a new research area in my opinion.

\vspace{.2cm}
$\bullet$ \hspace{.4cm}
In ref.\cite{Dobrescu1995}, Dobrescu proposed a supersymmetric TC model.
In this model, the mass hierarchy between the fermion generations arises
naturally. Furthermore, this model predicts
the CP asymmetries in B meson decays and in $\Delta S=1$ transitions to be
smaller by two orders of magnitude than the ones predicted in the SM.
Incorporating the supersymmetry into TC models is a novel and (perhaps)
very brave idea, further studies along this direction needed.

Stop here!  we are unable to list all new models proposed recently
in this paper. But one can understand from this short list that
the TC and ETC theories are now experiencing a somewhat rapid developing
period, after 10 years ``slow walking''.

Unfortunately, at present no ``standard'' or ``realistic''
(in its exact meaning) TC/ETC models which could resolve the basic
difficulties for the theories of DESB elegantly have
been emerged. But many progress have been achieved both in the model
construction and in the phenomenological analysis in recent years.
I think that we now begin ``running'' in the right direction, and
all these progress are valuable and indispensable
for the future success.

\vspace{.5cm}
\noindent {\bf ACKNOWLEDGMENT}

First of all, Z.J.Xiao would like to thank
Professor C.H.Chang and Professor Y.P.Kuang for their kind invitation.
It is pleasure for me to thank CCAST and the Seminar organizers
for their hospitality and
support. We thank Professor G.R.Lu, Professor Y.B.Ding and Professor X.Q.Li
for valuable discussions. This work was done partly during my stay in CCAST.
Z.J.Xiao acknowledge the support
of a Henan Province Outstanding  Teacher Foundation.
This work was supported in part by the National Natural
Science Foundation of China, and by the funds from
Henan Science and Technology  Committee.

\vspace{1cm}
\begin{center}
{\bf Figure Captions}
\end{center}
\begin{description}

\item[Fig.1:] The SM predictions for the ratio $R_b$ compared with the
current LEP data. The lower curve with solid triangle symbols
is the $R_b$ in SM at one-loop level,
while the upper line with solid square symbols represents the $R_b$ with
the inclusion of two-loop $0(\alpha \alpha_s)$ QCD contributions.
The upper error band corresponds to the current
data $R_b=0.2204\pm 0.0020$. The lower
error bar shows the current experimental measurement of $m_t$:
$m_t=180\pm 12\; GeV$.

\item[Fig.2:]  The constraints on the masses of the charged PGBs in the
QCD-like one-generation TC model for
$m_t=180\,GeV$. For more details see the text.

\end{description}

\end{document}